\documentclass[10pt,twocolumn,twoside]{IEEEtran}
\usepackage{amsmath}
\usepackage{amssymb}
\usepackage{algorithmicx}
\usepackage{algpseudocode}
\usepackage{graphicx}
\usepackage{subfigure}
\usepackage{epstopdf}
\usepackage{cite}
\usepackage{textcomp}
\usepackage{mathrsfs}
\usepackage{color}
\usepackage{booktabs}
\usepackage{cases}
\usepackage{setspace}
\usepackage{bm}
\usepackage{cuted}
\usepackage{booktabs}
\usepackage{stfloats}
\usepackage{makecell}
\usepackage{mathtools}
\usepackage[ruled,linesnumbered]{algorithm2e}

\makeatletter

\renewcommand{\maketag@@@}[1]{\hbox{\m@th\normalsize\normalfont#1}}%

\makeatother

\begin{document}
	
	\title{\huge Fairness-Aware Computation Offloading in Wireless-Powered MEC Systems with Cooperative Energy Recycling} 
	
	\author{ 
				Haohao~Qin, Bowen~Gu, Dong~Li, Xianhua~Yu, Liejun~Wang, \\ Yuanwei~Liu, \textit{Fellow, IEEE}, and Sumei~Sun, \textit{Fellow, IEEE} 
				\IEEEcompsocitemizethanks{			
				\IEEEcompsocthanksitem H. Qin, B. Gu, and L. Wang are with the School of Computer Science and Technology, Xinjiang University, Urumqi, Xinjiang 830049, China, and with Xinjiang Multimodal Intelligent Processing and Information Security Engineering Technology Research Center, Urumqi, Xinjiang 830049, China (e-mails: hhqin@stu.xju.edu.cn, bwgu@xju.edu.cn, wljxju@xju.edu.cn).
                \IEEEcompsocthanksitem D. Li is the School of Computer Science and Engineering, Macau University of Science and Technology, Macau 999078, China (e-mail:  dli@must.edu.mo).
                \IEEEcompsocthanksitem X. Yu is School of Electrical Engineering and Intelligentization, Dongguan University of Technology, Dongguan, China (e-mail: xianhuacn@foxmail.com).
                \IEEEcompsocthanksitem Y. Liu is with the Department of Electrical and Electronic Engineering, The University of Hong Kong, Hong Kong, China (e-mail: yuanwei@hku.hk).
                \IEEEcompsocthanksitem S. Sun is with the Institute of Infocomm Research, Agency for Science, Technology and Research, Singapore 138632 (e-mail: sunsm@ i2r.astar.edu.sg).
                    } \vspace{-10pt}	
	
	}



	\maketitle
	\thispagestyle{empty}
	\pagestyle{empty}
	
	\begin{abstract}

    In this paper, cooperative energy recycling (CER) is investigated in wireless-powered mobile edge computing (MEC) systems. Unlike conventional architectures that rely solely on a dedicated power source, wireless sensors are additionally enabled to recycle energy from peer transmissions. To evaluate system performance, a joint computation optimization problem is formulated that integrates local computing and computation offloading, under an $\alpha$-fairness objective that balances total computable data and user fairness while satisfying energy, latency, and task size constraints. Due to the inherent non-convexity introduced by coupled resource variables and fairness regularization,  a variable-substitution technique is employed to transform the problem into a convex structure, which is then efficiently solved using Lagrangian duality and alternating optimization. To characterize the fairness–efficiency tradeoff,  closed-form solutions are further derived for three representative regimes, i.e., zero fairness, common fairness, and max-min fairness, each offering distinct system-level insights. Numerical results validate the effectiveness of the proposed CER-enabled framework, demonstrating significant gains in throughput and adaptability over benchmark schemes, while the tunable $\alpha$ fairness mechanism provides flexible control over performance-fairness trade-offs across diverse scenarios.

	\end{abstract}
	\IEEEpeerreviewmaketitle
	\section{Introduction}
	The rapid proliferation of the Internet of Things (IoT) is reshaping modern society by enabling ubiquitous connectivity and intelligent data-driven services \cite{iotbackground}. Fueled by the rapid growth of vertical industries such as smart homes, industrial automation, intelligent transportation, and healthcare, IoT technologies are becoming deeply integrated into every aspect of daily life, driving a new era of pervasive digital transformation.  According to recent industry reports, the number of connected IoT devices is projected to surpass 41 billion by 2030 \cite{iotnumber}, generating over 80 zettabytes of data annually \cite{iotdata}.  This explosive growth imposes stringent demands for real-time data processing with ultralow latency and high reliability.
    
    However, conventional cloud computing architectures often fall short in meeting the stringent demands of latency, bandwidth, and privacy posed by delay-sensitive and task-intensive IoT applications. To address these challenges, mobile edge computing (MEC), also known as multi-access edge computing, has emerged as a promising paradigm by enabling the offloading of computation-intensive tasks from resource-constrained IoT devices to nearby edge servers, thus significantly reducing communication latency and improving real-time responsiveness \cite{9363323}. Despite that, MEC systems face a critical limitation: the restricted energy availability of IoT devices, which typically depend on embedded batteries \cite{10286271}. This limited energy capacity is insufficient to sustain continuous operation, especially in data-intensive or long-duration applications, thereby compromising the scalability and reliability of MEC-enabled networks and highlighting the need for sustainable energy provisioning tailored to their unique demands.
    
    Against this backdrop, wireless-powered communication networks (WPCNs) have garnered increasing attention as a sustainable solution for powering edge devices with limited energy resources \cite{9743350, 10589561}. By harnessing radio frequency (RF) energy transfer, WPCNs enable IoT devices to harvest energy from dedicated power sources (PSs) or ambient signals, thereby achieving battery-free and perpetual operation \cite{10736560}. This paradigm eliminates the need for manual battery replacement or recharging, significantly enhancing the scalability and maintainability of large-scale IoT deployments. Consequently,  the integration of WPCN and MEC technologies has given rise to a new paradigm, namely wireless-powered MEC systems, that address the critical energy bottleneck at the edge. By supporting continuous and battery-free operation of IoT devices, this architecture enhances the overall sustainability, scalability, and deployment flexibility of next-generation edge computing networks, thus attracting growing interest from both academia and industry.

	\subsection{Related Works}
	
	To fully unlock the potential of wireless-powered MEC systems, recent research efforts have increasingly concentrated on the joint optimization of energy utility, task offloading, and communication scheduling under stringent system constraints. For instance, in \cite{8960510}, a single-user wireless-powered MEC model was investigated, and an optimal scheme for energy allocation and partial task offloading was proposed. Building on this, \cite{9698985} introduced a mixed-offloading strategy that integrates both partial and binary modes, yielding improved transmission performance in high signal-to-noise ratio (SNR) regimes. In a multi-antenna setting, \cite{10239257} leveraged intelligent reflecting surfaces (IRS) to jointly optimize time allocation, access point selection, beamforming, and offloading decisions to maximize the computation rate.  Extending this framework, \cite{9887822} proposed a dynamic IRS beamforming approach applicable to both time-division multiple access (TDMA) and non-orthogonal multiple access (NOMA) schemes, thereby enhancing the computational throughput.  Furthermore, \cite{10210080} incorporated full-duplex technology in IRS-assisted systems,  enabling concurrent energy harvesting (EH) and data offloading to mitigate the inherent trade-off between power transfer and information transmission.

   Despite considerable advances in boosting computational throughput, comparatively limited attention has been devoted to optimizing energy-centric performance metrics. To bridge this gap, several works have investigated energy optimization from diverse perspectives. For instance, \cite{8234686} considered a multi-user wireless-powered MEC network and proposed a strategy to minimize the overall system energy consumption while satisfying latency requirements. Moving beyond absolute energy reduction, more recent research has shifted toward maximizing computational energy efficiency (CEE), seeking an optimal balance between throughput and energy expenditure. In \cite{9312671}, a NOMA-enabled MEC framework was studied by jointly optimizing the computation frequency and task execution time to improve CEE. Further advancing this direction, \cite{10039498, 9422161, 10301686, 10376310} integrated backscatter communication into wireless-powered MEC architectures, enabling partial task offloading via passive signal reflection and thereby substantially improving CEE under stringent energy constraints.
     
   Beyond enhancing energy usage, latency-sensitive execution remains a critical concern in wireless-powered MEC systems. On one hand, prolonged EH durations may cause devices to exceed their task deadlines before computation is completed; on the other, task backlogs at the MEC server can lead to inefficient utilization of computational resources. To mitigate these issues, \cite{9468700} proposed a wireless-powered MEC framework employing nonlinear rectifiers, in which task latency is minimized through the joint optimization of the rectifier’s power-splitting ratio, offloading power, local computing frequency, and offloading ratio. Similarly, \cite{10538223} investigated a NOMA-enabled wireless-powered MEC system supporting both binary and partial offloading modes, where the joint optimization of communication and computation strategies was performed to minimize the total task completion time. 
   
   To further improve system performance and reliability of wireless-powered MEC systems, recent efforts have incorporated advanced enabling technologies. For instance, \cite{10107791} employed IRS to improve wireless channel conditions, thereby boosting both EH efficiency and the security of computation offloading. In a similar vein,  \cite{10015648}  leveraged interference signals generated by dedicated energy beacons to thwart potential eavesdroppers, thus ensuring secure data transmission. To improve deployment flexibility and service coverage, unmanned aerial vehicles (UAVs) have also been introduced as mobile energy transmitters or computing platforms, dynamically adjusting their trajectories to better provision energy and offloading opportunities for ground devices \cite{9919310, 10097286}.  Extending this direction, \cite{10444003} investigated CEE fairness in UAV-assisted MEC systems with hybrid passive and active transmissions. Meanwhile, \cite{10050170} explored device-to-device cooperation, where idle user nodes assist in task execution, enabling more distributed and resource-efficient offloading. Moreover, \cite{10756636} investigated multi-access-point wireless-powered MEC architectures, wherein user devices can harvest energy from multiple sources and dynamically select optimal offloading destinations in response to time-varying network conditions.

	\subsection{Motivations and Contributions}

    Despite the promising potential of wireless-powered MEC in enabling sustainable and battery-free IoT deployments, most existing architectures still adopt a conventional energy supply mode, in which devices passively harvest power from dedicated RF sources and are treated as isolated energy recipients. This rigid and non-cooperative paradigm often results in severe energy imbalance, especially in dense networks where some nodes accumulate surplus energy while others suffer acute shortages. Moreover, the static nature of such provisioning means that performance improvements typically require increasing the overall energy budget, which contradicts the green conception of IoT systems. In contrast, energy recycling (ER) introduces a paradigm shift by enabling devices to harvest and recycle energy from peer transmissions, unlocking “free” energy that would otherwise be wasted and enhancing network-wide energy utilization.  Nevertheless, this concept has received only limited exploration in the WPCN systems, with only a handful of studies making initial attempts. For example, \cite{9888066} proposed a backscatter-assisted ER scheme that enables passive tags to scavenge energy from ambient transmissions, while \cite{10083178} designed an active ER framework in IRS-assisted networks to enable more flexible and targeted energy exchanges.

    Although ER demonstrates considerable promise, most prior work has predominantly concentrated on its physical-layer energy transfer aspects, leaving the computation-oriented challenges in MEC systems largely unaddressed. In contrast to conventional WPCNs, which mainly focus on ensuring sufficient energy for data transmission, wireless-powered MEC networks require a holistic orchestration of both energy and computing resources, where task partitioning and offloading must be jointly optimized to balance performance and fairness. This gap motivates a critical question: Is it possible to design a wireless-powered MEC architecture that fully harnesses ER to maximize energy utilization while ensuring computational fairness among heterogeneous users?  To answer this question, we develop a fairness-aware cooperative energy recycling (CER) protocol that seamlessly integrates energy cooperation with the unified design of local computing and computation offloading, thereby unlocking the synergy between communication and computation processes.

    In a nutshell, the main contributions of this paper can be summarized as follows.
	
	\begin{itemize}
		\item To evaluate the proposed CER protocol, we design a joint resource allocation framework that captures the intrinsic coupling between communication and computation resources, while incorporating energy causality, latency, and power constraints, to maximize the total computable data of all users under an $\alpha$-fairness criterion.
		\item To tackle this challenging problem, we first adopt maximal ratio combining (MRC) to simplify the receive beamforming design, followed by appropriate variable substitutions to reformulate the problem into a convex form, thereby enabling the application of standard convex optimization techniques to obtain the optimal solution efficiently.   
		\item To further explore the fairness–efficiency tradeoff in the proposed CER framework, we investigate three representative $\alpha$-fairness regimes, namely, zero fairness ($\alpha=0$), common fairness ($0 < \alpha < \infty$), and max-min fairness ($\alpha \rightarrow \infty$), and derive closed-form solutions for each using dual decomposition and alternating optimization, thereby elucidating fairness-induced performance variations. In addition, we quantify the performance gains achieved by CER under specific configurations.
		\item Simulation results validate the good convergence of the proposed algorithms across all three fairness regimes, reveal the inherent fairness–performance tradeoff under varying $\alpha$, and demonstrate, through comparisons with benchmark schemes, the superiority of the proposed approach in maximizing the total computable data size.
	\end{itemize}

    	\subsection{Organization and Notations}
	
	The remainder of this paper is organized as follows. Section II presents the system model, including the energy supply and computation offloading mechanisms. Section III formulates the $\alpha$-fairness optimization problem and develops an efficient solution framework, with three special cases analyzed for further insight. Section IV discusses the simulation results, and Section V concludes the paper.

	\section{System Model}
	
	\begin{figure*}[t]
		\centerline{\includegraphics[width=4.0in]{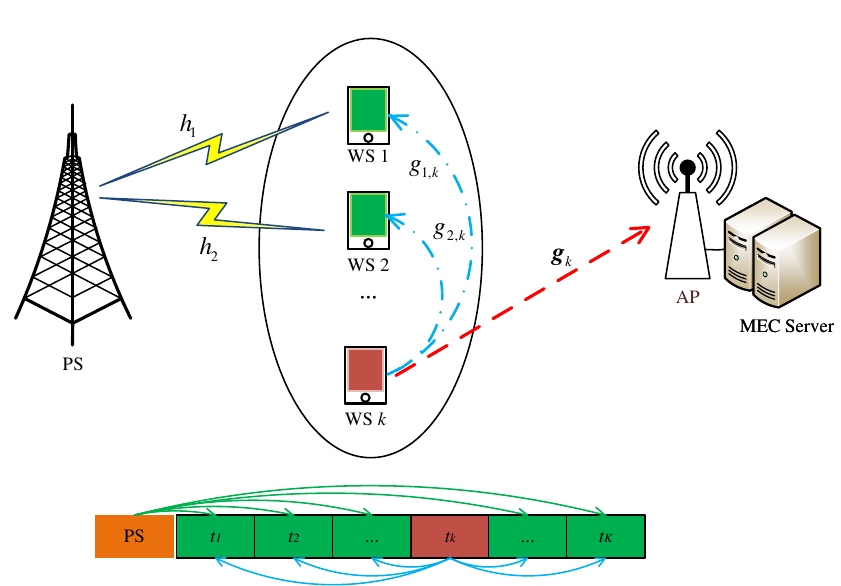}}
		\caption{A WPCN-assisted MEC system with energy recycling.}
		\label{fig1}
	\end{figure*} 
	
	\subsection{System Structure and Transmission Mechanism}
    
	   We consider a WPCN-assisted MEC system, as illustrated in Fig. \ref{fig1}, consisting of a single-antenna power station (PS), $K$ single-antenna wireless sensors (WSs), indexed by $k\in\mathcal{K}=\{1,2,\cdots, K\}$, and a multi-antenna access point (AP) equipped with $N$ receive antennas, indexed by $n\in\mathcal{N}=\{1,2,\cdots, N\}$.  An MEC server is co-located with the AP to support remote task execution. Each WS is assumed to have specific computational tasks and is capable of harvesting energy to replenish the power consumed during task execution, regardless of whether the tasks are processed locally or offloaded to the MEC server.
	   
      We adopt a quasi-static flat-fading channel model, in which the channel coefficients remain constant within each time slot but vary independently across different slots \cite{9887822}.  The system operates in a frame-based manner, where each frame of duration $T$ consists of two phases: a computation offloading phase and an edge computing phase, designed to satisfy the latency constraints of computational tasks. The execution time for edge computing and result downloading is considered negligible \cite{10015648} due to the MEC server’s high computational capability and the small size of computation results (i.e., $\epsilon\approx 0$). For task execution, each WS performing local computing can utilize the entire duration $T$, while a TDMA-based scheme is employed for computation offloading, in which each WS is assigned an exclusive time slot $t_k$ for data transmission to avoid mutual interference. Therefore, the total transmission time must satisfy
	   \begin{equation} \label{t1}
	   	\sum_{k=1}^K t_k \le T-\epsilon.\\
	   \end{equation}

	\subsection{ Energy Supply Mechanism}
	
    Since single-antenna WSs cannot perform EH and data transmission at the same time, they harvest energy from the PS during their non-transmission slots \cite{10107791}.
    Moreover, they can recycle energy from the signals transmitted by other WSs during their respective transmission slots. The received signal at the $k$-th WS for EH can therefore be expressed as
	\begin{equation} \label{s1}
		y_{k}^{\text{Rx}}=\sum_{i=1,i\neq k}^K \sqrt{P_i} h_ks_i+\sum_{i=1,i\neq k}^K \sqrt{p_i} g_{i,k} x_i+ n_k,
	\end{equation}
where $h_k$ represents the channel coefficients between the PS and the $k$-th WS, and $g_{i,k}$ denotes the channel coefficients between the $i$-th WS and the $k$-th WS. $s_i$ and $x_i$ are the transmitted symbols from the PS and the $i$-th WS, respectively, satisfying $\mathbb{E}(|s_i|^2)=1$ and $\mathbb{E}(|x_i|^2)=1$. $n_k$ is the additive noise at the $k$-th WS. Furthermore, $P_i$ and $p_i$ represent the transmit powers of the PS and the $i$-th WS, respectively, during the $i$-th slot. 
	
The contribution of the noise to the harvested energy is considered negligible, as the noise power is typically several orders of magnitude lower than the received signal power \cite{10376310}.  Therefore, the total energy harvested by the $k$-th WS is expressed as
\begin{equation} \label{eh1} E_k^{\text{EH}}=\underbrace{\sum\limits_{i=1,i\neq k}^K \eta t_i P_i |h_k|^2}_{\text{Harvested from the PS}} + \underbrace{\sum\limits_{i=1,i\neq k}^K \eta t_i p_i |g_{i,k}|^2}_{\text{Recycled from other WSs}}, 
\end{equation} 
where $t_i$ denotes the transmission time allocated to the $i$-th WS, and $\eta \in (0, 1]$ denotes the efficiency of energy conversion. 

\subsection{Task Execution Mechanism}
It is assumed that computational tasks are bitwise independent, supporting arbitrary partitioning of task data \cite{10486847}. Here, a partial offloading mechanism is considered, enabling the simultaneous execution of local computation and remote offloading.

\subsubsection{Local Computing} Let $C_k$ represent the number of central processing unit (CPU) cycles required to compute one bit of data. The number of bits computed locally by the $k$-th WS can be expressed as
\begin{equation} \label{lc1}
	R_k^{\text{LC}}=\frac{T f_k}{C_k},
\end{equation}
where $f_k$ denotes the CPU operating frequency of the $k$-th WS.

Each WS adopts an advanced dynamic voltage and frequency scaling (DVFS) technique \cite{10376310}. For analytical tractability, it is assumed that the CPU frequency $f_k$ remains fixed during each operational frame. Thus, the energy consumption for local computing at the $k$-th WS is given by
\begin{equation} \label{lc2}
	E_k^{\text{LC}}= T\phi_kf_k^3,
\end{equation}
where $\phi_k$ denotes the switched capacitance coefficient of the $k$-th WS.	

\subsubsection{Computation Offloading}  The signal received at the AP from the $k$-th WS during computation offloading is expressed as
\begin{equation} \label{s2}
	y_k^{\text{Tx}}=\bm w_k^{\text{H}} \bm g_k \sqrt{p_k} x_k+ \bm w_k^{\text{H}} \bm z_{\text{M}},
\end{equation}
where $\bm g_k \in \mathbb{C}^{N\times1}$ denotes the channel vector between the $k$-th WS and the MEC server, and $\bm z_{\text{M}}$ represents the noise at the MEC server, with $\bm z_{\text{M}} \sim \mathcal{CN}(0, \delta_{\text{M}}^2\bm I_N)$. Additionally, $\bm w_k \in \mathbb{C}^{N\times1}$ is the receive beamforming vector, satisfying $||\bm w_k||^2=1$.

Consequently, the data size offloaded by the $k$-th WS can be expressed as 
\begin{equation} \label{co1}
	R_k^{\text{CO}}=t_k B \log_2\left( 1+\dfrac{p_k|\bm w_k^{\text{H}}\bm g_k|^2}{\delta_{\text{M}}^2}\right) ,
\end{equation}
where $B$ denotes the system bandwidth allocated for computation offloading. 
	
Therefore, the total data size processed by the $k$-th WS, combining local computing and offloading, is given by
\begin{equation} \label{rt}
	R_k=R_k^{\text{LC}}+R_k^{\text{CO}}.
\end{equation}	

Moreover, the energy needed by the $k$-th WS for its task can be given by
\begin{equation} \label{et}
	E_k^{\text{EC}}=E_k^{\text{LC}}+E_k^{\text{CO}},
\end{equation}
where $E_k^{\text{CO}}=p_kt_k$ denotes the energy consumed by the $k$-th WS for computation offloading.

	\section{Problem Formulation and Algorithm Design}
	
	\subsection{Problem Formulation}

    In wireless-powered MEC networks, the coexistence of diverse computational demands and asymmetric energy availability among WSs naturally raises concerns about service disparity. Without proper coordination, weaker WSs may suffer from insufficient computing opportunities, which not only degrades their quality of experience but also leads to unbalanced overall system performance. To address this issue and promote fairness across multiple WSs, we adopt the $\alpha$-fairness criterion. Specifically, the $\alpha$-fair utility function\cite{9625847} is defined as
	     \begin{equation}\label{f1}
	     	{u_\alpha }(R_k) = \left\{ {\begin{array}{*{20}{c}}
	     			{\ln (R_k),{\rm{ \;  if\; }}\alpha  = 1}, {\qquad}{\quad}\\
	     			{\dfrac{1}{1 - \alpha } (R_k)^{1 - \alpha },{\rm{ \; if \;}}\alpha  \ne 1,\alpha  \ge 0},  \\
	     	\end{array}} \right.
	     \end{equation}
	     where $\alpha$ is a fairness control parameter that tunes the trade-off between maximizing overall throughput and enhancing fairness. Based on this criterion, the joint resource allocation problem can be formulated as follows, i.e.,
		\begin{equation}  \label{p1}  
	\begin{aligned}	
		& \underset{P_k, t_k, p_k, f_k, \bm w_k}{\mathop{\max }}\,  {\sum\limits_{k=1}^K  {u_\alpha }(R_k)}\\
		&\text{s.t.}~
		C_1: P_k \le P_{\max},  \forall k, \\
		&\quad ~~C_2: \sum_{k=1}^K t_k \le T-\epsilon, \\
		&\quad ~~C_3: E_k^{\text{EC}}\le E_k^{\text{EH}}, \forall k, \\
		&\quad ~~C_4: f_k \le f_k^{\max}, \forall k, \\
		&\quad ~~C_5: ||\bm w_k||^2=1, \forall k,\\
	    &\quad ~~C_6:  R_k \ge R_k^{\min}, \forall k,  \\
	\end{aligned}
\end{equation}  
 where $P_{\max}$ denotes the maximum power supported by the PS, $f_k^{\max}$ is the maximum CPU frequency of the $k$-th WS, and $R_k^{\min}$ represents the minimum data size that the $k$-th WS must compute. In problem (\ref{p1}), $C_1$ ensures that the transmit power of the PS does not exceed the hardware power limit. $C_2$ guarantees that the total transmission time remains within the frame duration after accounting for computation latency. $C_3$ enforces the energy causality constraint, ensuring that the energy consumption for computation and offloading does not exceed the harvested energy. $C_4$ limits the CPU frequency of each WS to its maximum supported value, adhering to device capabilities. $C_5$ normalizes the receive beamforming vectors to ensure unit-norm processing at the AP. $C_6$ ensures that each WS processes at least the minimum required amount of data to meet user demands.
 
\textbf{\textit{ Remark 1:}} 
 Based on (\ref{f1}), the parameter $\alpha$ directly determines the balance between throughput maximization and fairness enhancement.  Specifically, $u_\alpha(R_k)$ corresponds to zero fairness when $\alpha=0$, proportional fairness when $\alpha=1$, harmonic mean fairness when $\alpha=2$, and approaches max–min fairness as $\alpha \to +\infty$ \cite{8294215}.
 When $\alpha=0$, the objective reduces to maximizing the total computable data without fairness considerations. As $\alpha$ increases, the allocation becomes more equitable, eventually converging to the max–min fairness regime. This flexibility enables system designers to tailor resource allocation strategies to diverse QoS and sustainability requirements.

\subsection{Problem Transformation}

As can be seen, the inherent non-convexity of problem (\ref{p1}) stems from the strong coupling among transmission times $t_k$, transmit powers $p_k$ and $P_k$, and receive beamforming vectors $\bm{w}_k$, which poses significant challenges for deriving the optimal solution. To facilitate tractable analysis, we adopt MRC, which decouples the beamforming design by independently maximizing the received SNR for each WS without affecting the other variables. Consequently, the optimal receive beamforming vector for the $k$-th WS is given by
\begin{equation} \label{m7}
	\bm w_n\triangleq \frac{\bm g_k}{||\bm g_k||}.\\
\end{equation}	

Substituting (\ref{m7}) into (\ref{rt}), the computable data size for the $k$-th WS can be rewritten as
\begin{equation}
	\hat R_k =R_k^{\text{LC}}+t_kB\log_2\left( 1+\dfrac{p_k||\bm g_k||^2}{\delta_{\text{M}}^2}\right).
\end{equation}

While the dependency on $\bm{w}_k$ has been removed, problem (\ref{p1}) remains challenging due to the residual coupling between $t_k$, $p_k$, and $P_k$. To decouple these variables and further simplify the problem structure, we introduce the substitutions $\bar{p}_k = p_k t_k$ and $\bar{P}_k = P_k t_k$, representing the transmission energies for computation offloading and EH in their respective time slots. Accordingly, problem (\ref{p1}) can be reformulated into a more tractable form, which is detailed as follows, i.e.,
		\begin{equation}  \label{p3}  
	\begin{aligned}	
		& \underset{\bar P_k, t_k, \bar p_k, f_k}{\mathop{\max }}\,  \sum\limits_{k=1}^K  {u_\alpha }(\bar R_k)\\
		&\text{s.t.}~C_2, C_4,\bar C_1: \bar P_k \le P_{\max}t_k, \forall k,\\
		&\quad~~ \bar C_3: \bar E_k^{\text{EC}}\le \bar E_k^{\text{EH}}, \forall k,\\
		&\quad ~~\bar C_6: \bar R_k \ge R_k^{\min}, \forall k,  \\
		&\quad ~~C_7: \bar P_k\ge 0, \bar p_k\ge 0, \forall k,
	\end{aligned}
\end{equation}
where 
\begin{equation}
	\bar R_k =R_k^{\text{LC}}+t_kB\log_2\left( 1+\dfrac{\bar p_k||\bm g_k||^2}{t_k\delta_{\text{M}}^2}\right),
\end{equation}
\begin{equation}
	\bar E_k^{\text{EC}}=T\phi_kf_k^3+\bar p_k,
\end{equation}
\begin{equation}
	\bar E_k^{\text{EH}}=\sum\limits_{i=1,1\neq k}^K \eta \bar P_i |h_k|^2 + \sum\limits_{i=1,1\neq k}^K \eta \bar p_i |g_{i,k}|^2.
\end{equation}

Compared with problem (\ref{p1}), problem (\ref{p3}) attains a significantly more tractable structure, with key nonconvex couplings eliminated and all constraints expressed in convex or convex-like forms, thereby facilitating efficient algorithm design. Building on this reformulation, we next examine the convexity structure of problem (\ref{p3}), focusing on whether the objective is concave and the feasible region is convex, which are prerequisites for applying convex optimization frameworks. This leads to the following theorem:

\textbf{\textit{Theorem 1:}} The objective function of problem (\ref{p3}) is concave with respect to the associated optimization variables.

\textit{Proof:} Please see Appendix A.

\subsection{Algorithm Design}

Before solving problem (\ref{p3}), we first present the following lemma.

\textbf{\textit{Lemma 1:}} The optimal $t_k^*$ of problem (\ref{p3}) must satisfy 
\begin{equation} \label{l1}
	\sum_{k=1}^K t_k^* = T-\epsilon.\\
\end{equation}

\textit{Proof:} Please see Appendix B.

\textbf{\textit{Remark 2:}} Lemma 1 implies that, at optimality, the entire latency budget is fully utilized for computation offloading and local processing. Any idle time could be reassigned to these tasks to strictly improve the objective value. This reflects the intuitive fact that extending transmission time increases WSs’ achievable rates, and it also simplifies the optimization by fixing the total time allocation.

Since all the constraints in problem (\ref{p3}) are either affine or concave in the optimization variables, and the objective function $\sum_{k=1}^K u_\alpha(\bar R_k)$ is concave, problem (\ref{p3}) is convex. Therefore, standard convex optimization methods, such as the interior-point algorithm or CVX \cite{Boyd_Vandenberghe_2004}, can be employed to efficiently obtain the global optimum ${\bar P_k^*, t_k^*, \bar p_k^*, f_k^*}$. The optimal solution to the original problem (\ref{p1}) is then recovered via
\begin{equation} \label{os} 
	t_k = t_k^*, \quad f_k = f_k^*, \quad P_k = \frac{\bar P_k^*}{t_k^*}, \quad p_k = \frac{\bar p_k^*}{t_k^*}. 
\end{equation} 
Thus, a complete solution to the original problem (\ref{p1}) is achieved.

\subsection{Case Analysis}

To further elucidate how the fairness control parameter $\alpha$ influences the optimal resource allocation and overall system performance, we study three representative cases, i.e., zero fairness ($\alpha=0$), common fairness ($0<\alpha<+\infty$), and max-min fairness ($\alpha \to +\infty$), and analyze their distinct implications.

\subsubsection{Zero fairness}
When $\alpha=0$, problem (\ref{p3}) reduces to the zero-fairness case, formulated as
 \begin{equation}  \label{p31}  
 	\begin{aligned}	
 	\textbf{P1}:	& \underset{\bar P_k, t_k, \bar p_k, f_k}{\mathop{\max }}\,  \sum\limits_{k=1}^K   \bar R_k\\
 		&\text{s.t.}~\bar C_1,C_2,\bar C_3, C_4,  \bar C_6, C_7. \\
 	\end{aligned}
 \end{equation}
Although (\ref{p31}) is convex, a closed-form global optimum is still difficult to obtain due to the residual coupling among the decision variables. To address this, we adopt an alternating optimization approach. Specifically, given fixed transmit powers at the PS and WSs, we jointly optimize the transmission time and CPU frequency; then, with the optimized transmission time and CPU frequency fixed, we update the transmit powers. This decoupling ensures that each subproblem admits a closed-form solution. The overall procedure is summarized in the proposed zero-fairness-based algorithm (ZFBA) in \textbf{Algorithm 1}, and the closed-form solutions for both subproblems are derived below.

\textbf{\textit{Theorem 2:}} 
With the given $\bar P_k$ and $\bar p_k$, the optimal transmission time and CPU frequency for problem (\ref{p31}) can be given by
\begin{equation} \label{ot}
	t_k^*= \dfrac{\bar p_k||\bm g_k||^2}{\delta_{\text{M}}^2f_t^{-1}\left( \dfrac{\omega_2-\omega_1^kP_{\max}}{(1+\omega_5^k)B}\right)}, \forall k,
\end{equation}
\begin{equation} \label{of}
	f_k^*=\sqrt{\dfrac{(1+\omega_5^k)\frac{ T}{C_k}-\omega_4^k}{3\omega_3^k T\phi_k}}, \forall k,
\end{equation}
where $f_t^{-1}(x)$ is the inverse function of $f_t(x)$, $f_t(x)$ and $f_t^{-1}(x)$ are defined as (\ref{ap22}) and (\ref{apc4a}), respectively. Besides,   $\omega_1^k$, $\omega_3^k$, $\omega_4^k$, and $\omega_5^k$ represent non-negative Lagrange multipliers associated with the corresponding constraints of problem (\ref{p31}), and $\omega_2 > 0$ follows from Lemma 1. 

Conversely, with the given $t_k$, $f_k$, the optimal $\bar p_k$ and $\bar P_k$ are given by
\begin{equation} \label{op}
	\bar p_k^*=\left[ \frac{(1+\mu_3^k)Bt_k}{\ln 2(\psi+\mu_2^k-\mu_5^k)}-\frac {t_k\delta_{\text{M}}^2}{||\bm g_k||^2}\right]^+, \forall k, 
\end{equation}
\begin{equation} \label{obp}
		\bar P_k^*= [P_{\max}t_k]^+,  \forall k,
\end{equation}
where $\mu_2^k$, $\mu_3^k$, and $\mu_5^k$ represent non-negative Lagrange multipliers associated with the corresponding constraints of problem (\ref{p31}), and $[x]^+=\max\{x,0\}$. 

\textit{Proof:} Please see Appendix C.

\textbf{\textit{Remark 3:}} In the zero-fairness regime, the system aims to maximize aggregate throughput without explicitly considering fairness among WSs. 
As shown in (\ref{ot}),  $t_k^*$ is directly influenced by both the transmit power
$p_k$ and the channel gain $||\bm g_k||^2$, resulting in a strongly biased allocation that favors high-quality links. This allocation mechanism intensifies disparities among devices, as the logarithmic rate function accentuates performance gaps caused by channel variations.
Meanwhile, (\ref{of}) reveals that $f_k^*$ increases with the frame duration $T$, encouraging higher local computation loads when longer computation windows are available.
Additionally, (\ref{op}) highlights that $\bar p_k^*$ adopts a threshold-based activation rule, whereby a WS engages in offloading only if its channel gain and marginal utility exceed the composite “energy–fairness price” reflected in the dual variables; otherwise, it defaults to local execution.  Moreover, (\ref{obp}) shows that the PS always operates at its maximum power $P_{\max}$ during each WS’s time slot to maximize throughput, leading to aggressive energy utilization. Although this approach achieves the highest overall throughput, it can lead to severe performance starvation for weak-channel devices, making it most suitable for private MEC systems or scenarios where maximizing total system efficiency outweighs fairness considerations.

\begin{algorithm}[t]
	\caption{Zero Fairness-Based Algorithm }
	\SetAlgoLined
	\KwIn{System parameters: $P_{\max}$, $T$, $\epsilon$, $f_k^{\max}$, $B$, $\delta_{\text{M}}^2$, etc.}
	\KwOut{Optimal solution $\{\bar P_k^*, t_k^*, \bar p_k^*, f_k^*\}$.}
	
	Initialize $\{\bar P_k, t_k, \bar p_k, f_k\}$ with feasible values;\\
	
	\Repeat{convergence of $\{\bar P_k, t_k, \bar p_k, f_k\}$}{
		\textbf{Step 1:} Optimize $t_k$ and $f_k$ with fixed $\bar P_k$ and $\bar p_k$\\
		\quad Solve (\ref{ot}) and (\ref{of}) using the current Lagrange multipliers $\{\omega_1^k, \omega_2, \omega_3^k, \omega_4^k, \omega_5^k\}$.\\
		
		\textbf{Step 2:} Optimize $\bar p_k$ and $\bar P_k$ with fixed $t_k$ and $f_k$\\
		\quad Solve (\ref{op}) and (\ref{obp}) using updated multipliers $\{\mu_1^k, \mu_2^k, \mu_3^k, \mu_4^k, \mu_5^k\}$.
	}
	
	\textbf{Step 3:} Recover original variables according to (\ref{os}).\\
	\KwResult{Return $\{P_k^*, t_k^*, p_k^*, f_k^*\}$.}
\end{algorithm}		

\subsubsection{Common fairness} When $0<\alpha<+\infty$, the resource allocation problem explicitly embeds a tradeoff between system throughput and user fairness via the $\alpha$-fair utility function.
To decouple the nonlinear dependency of $\bar R_k$ in the utility, an auxiliary variable $\chi_k$ is introduced with $\bar R_k \geq \chi_k \geq R_k^{\min}$. Then, problem (\ref{p3}) is transformed into
		\begin{equation}  \label{p32a}  
	\begin{aligned}	
	\textbf{P2}:	& \underset{\bar P_k, t_k, \bar p_k, f_k, \chi}{\mathop{\max }}\,  \sum\limits_{k=1}^K  {u_\alpha }(\chi_k) \\
		&\text{s.t.}~\bar C_1,C_2,\bar C_3, C_4,  C_7,\\
		&\quad ~~ \hat C_6: \chi_k\ge R_k^{\min}, \forall k, \\
	   &\quad ~~ C_8: \bar R_k \ge \chi_k, \forall k. \\
	\end{aligned}
\end{equation}

Before solving problem (\ref{p32a}), we first present the following lemma.

\textbf{\textit{Lemma 2:}}  The optimal solution ($\bar P_k^*$, $t_k^*$, $\bar p_k^*$, $f_k^*$, $\chi_k^*$)
 of problem (\ref{p32a}) satisfies the following relationship, i.e., 
 \begin{equation} \label{la2}
 	\chi_k^*=\bar R_k (\bar P_k^*, t_k^*, \bar p_k^*, f_k^*), \forall k,
 \end{equation}
 and thus problem (\ref{p32a}) is equivalent to problem (\ref{p3}) in terms of optimal solutions.
 
 \textit{Proof:} Please see Appendix D.
 
\textbf{\textit{Theorem 3:}} With the given $\bar P_k$ and $\bar p_k$, the optimal $t_k$ and $f_k$ for problem (\ref{p32a}) can be given by
\begin{equation} \label{ot1}
	t_k^*= \dfrac{\bar p_k||\bm g_k||^2}{\delta_{\text{M}}^2f_t^{-1}\left( \dfrac{\zeta_2-\zeta_1^kP_{\max}}{\zeta_6^kB}\right)}, \forall k,
\end{equation}
\begin{equation} \label{of1}
	f_k^*=  \sqrt{\dfrac{\zeta_6^k\frac{ T}{C_k}-\zeta_4^k}{3\zeta_3^kT\phi_k}}, \forall k,
\end{equation}
where $\zeta_1^k$,  $\zeta_2$, $\zeta_3^k$, $\zeta_4^k$, and $\zeta_6^k$ represent non-negative Lagrange multipliers associated with the corresponding constraints of problem (\ref{p32a}).

Conversely, with the given $t_k$, $f_k$, the optimal $\bar P_k^*$ remains the same as in Case I, i.e., (\ref{obp}), and the optimal $\bar p_k$ for problem (\ref{p32a}) is given by
\begin{equation} \label{op1}
	\bar p_k^*= \left[ \frac{\theta_4^kBt_k}{\ln 2\left( {\theta_2^k-\theta_6^k}\right) }-\frac {t_k\delta_{\text{M}}^2}{||\bm g_k||^2}\right]^+, \forall k,
\end{equation}
where  $\theta_2$, $\theta_4^k$, and $\theta_6^k$ represent non-negative Lagrange multipliers associated with the corresponding constraints of problem (\ref{p32a}).

 \textit{Proof:} Please see Appendix E.

\textbf{\textit{Remark 4:}} In the common fairness regime, the system balances total throughput and user fairness through the $\alpha$-fair utility function. Compared with Case I, $t_k^*$ in (\ref{ot1}) is no longer solely driven by the channel gain $||\bm g_k||^2$ and the transmit power $\bar p_k$; instead, it is critically moderated by the multiplier $\zeta_6^k$, which redistributes the time budget $T-\epsilon$ and the desire to push $\bar R_k$ upwards.  Likewise, (\ref{of1}) indicates that $f_k^*$ is adjusted not only by energy constraints and frame duration but also by fairness considerations, allowing heterogeneous devices to balance their computation loads. 
Although (\ref{op1}) still indicates that $\bar p_k$ increases with the channel gain, the influence of $\theta_4^k$ mitigates excessive bias toward strong users. The fairness-adjusted threshold condition, i.e., $||\bm g_k||^2 \geq \frac{\ln 2 (\theta_2^k - \theta_6^k) t_k \delta_{\text{M}}^2}{\theta_4^k B t_k}$, ensures that offloading eligibility is also shaped by fairness constraints.  Moreover, as in Case I, the PS-side power $\bar P_k$ is pushed to its maximum, since boosting RF energy transfer benefits all users and does not undermine fairness. Overall, the common fairness design provides a principled compromise between efficiency and fairness, making it well-suited for public or shared MEC deployments where moderate fairness is desired. Nevertheless, if $\alpha$ (or the associated multipliers) is not sufficiently large, extremely weak users may still face non-negligible performance degradation.

\begin{algorithm}[t]
	\caption{Common Fairness-Based Algorithm}
	\SetAlgoLined
	\KwIn{System parameters: $P_{\max}$, $T$, $\epsilon$, $f_k^{\max}$, $B$, $\delta_{\text{M}}^2$, etc.}
	\KwOut{Optimal solution $\{\bar P_k^*, t_k^*, \bar p_k^*, f_k^*, \chi_k^*\}$.}
	
	Initialize $\{\bar P_k, t_k, \bar p_k, f_k, \chi_k\}$ with feasible values;\\
	
	\Repeat{convergence of $\{\bar P_k, t_k, \bar p_k, f_k, \chi_k\}$}{
		\textbf{Step 1:} Optimize $t_k$ and $f_k$ with fixed $\bar P_k$ and $\bar p_k$\\
		\quad Solve (\ref{ot1}) and (\ref{of1}) using the current Lagrange multipliers $\{\zeta_1^k, \zeta_2, \zeta_3^k, \zeta_4^k, \zeta_5^k, \zeta_6^k\}$.\\
		
		\textbf{Step 2:} Update slack variable $\chi_k$\\
		\quad Update $\chi_k = (\zeta_6^k - \zeta_5^k)^{-\frac{1}{\alpha}}$ according to (\ref{apd2a}).\\
		
		\textbf{Step 3:} Optimize $\bar p_k$ and $\bar P_k$ with fixed $t_k$ and $f_k$\\
		\quad Solve (\ref{op1}) and (\ref{obp}) using the updated Lagrange multipliers $\{\theta_1^k, \theta_2^k, \theta_3^k, \theta_4^k, \theta_5^k, \theta_6^k\}$.\\
		
		\textbf{Step 4:} Update slack variable $\chi_k$ again\\
		\quad Update $\chi_k = (\theta_4^k - \theta_3^k)^{-\frac{1}{\alpha}}$ according to (\ref{apd7b}).
	}
	
	\textbf{Step 5:} Recover original variables according to (\ref{os}).\\
	\KwResult{Return $\{P_k^*, t_k^*, p_k^*, f_k^*\}$.}
\end{algorithm}

The common fairness-based algorithm (CFBA) is summarized in \textbf{Algorithm 2}, which is similar to \textbf{Algorithm 1}. The main difference is that we need to update the slack variable $\chi_k$ according to (\ref{apd2a}) and (\ref{apd7b}) in each iteration until it converges.

\subsubsection{Max-min fairness} When $\alpha=+\infty$, the resource allocation objective shifts to maximizing the minimum utility across all WSs, thereby ensuring strict fairness among devices.
In this extreme case, problem (\ref{p3}) can be equivalently reformulated as the following max-min optimization problem, i.e., 
 \begin{equation}  \label{p33}  
	\begin{aligned}	
	\textbf{P3}:	& \underset{\bar P_k, t_k, \bar p_k, f_k}{\mathop{\max }}\, \underset{\forall k} {\mathop{\min }}\, \bar R_k \\
		&\text{s.t.}~\bar C_1,C_2,\bar C_3, C_4,  \bar C_6, C_7. \\
	\end{aligned}
\end{equation}

To make problem (\ref{p33}) more tractable, we introduce an auxiliary variable $\gamma$ such that, $\underset{\forall k} {\mathop{\min }}\, \bar R_k \ge \gamma$, thereby equivalently transforming problem (\ref{p33}) into
 \begin{equation}  \label{p33a}  
	\begin{aligned}	
		& \underset{\bar P_k, t_k, \bar p_k,  f_k, \gamma}{\mathop{\max }}\, \gamma  \\
		&\text{s.t.}~\bar C_1,C_2,\bar C_3, C_4, \bar C_6, C_7, \\
		&\quad ~~ C_8:  \bar R_k \ge \gamma, \forall k.
	\end{aligned}
\end{equation}

Before solving problem (\ref{p33a}), we first present the following lemma.

\textbf{\textit{Lemma 3:}} Problem (\ref{p33a}) ensures a tradeoff between the computable data size and energy consumption for the most disadvantaged WSs.
The optimal solution ${\bar P_k^*, t_k^*, \bar p_k^*, f_k^*, \gamma^*}$ satisfies
\begin{equation} \label{la3} \underset{\forall k}{\mathop{\min}} ~ \bar R_k(\bar P_k^*, t_k^*, \bar p_k^*, f_k^*)   = \gamma^*, \end{equation} and the optimal solution of problem (\ref{p33a}) is equivalent to that of problem (\ref{p33}).

 \textit{Proof:} The result follows directly from the argument in Appendix D, and a similar contradiction method can be applied here, so the proof is omitted for brevity.

\textbf{\textit{Theorem 4:}} With the given $\bar P_k$ and $\bar p_k$, the optimal $t_k$ and $f_k$ for problem (\ref{p33a}) can be given by
\begin{equation} \label{ot2}
	t_k^*= \dfrac{\bar p_k||\bm g_k||^2}{\delta_{\text{M}}^2f_t^{-1}\left( \dfrac{\lambda_2-\lambda_1^kP_{\max}}{(\lambda_5^k+\lambda_6^k) B}\right)}, \forall k,
\end{equation}
\begin{equation} \label{of2}
	f_k^*= \sqrt{\dfrac{(\lambda_5^k+\lambda_6^k)\frac{ T}{C_k}-\lambda_4^k}{3\lambda_3^k T\phi_k}},  \forall k,
\end{equation}
where $\lambda_1^k$, $\lambda_2$, $\lambda_3^k$, $\lambda_4^k$, $\lambda_5^k$, and $\lambda_6^k$ represent non-negative Lagrange multipliers associated with the corresponding constraints of problem (\ref{p33a}).

Conversely, with the given $t_k$, $f_k$, the optimal $\bar P_k^*$ also follows the same expression as in Case I, namely, (\ref{obp}), and the optimal $\bar p_k$ is given by
\begin{equation} \label{op2}
	\bar p_k^*= \left[ \dfrac{(\varepsilon_3^k+\varepsilon_6^k)Bt_k||\bm g_k||^2}{\ln 2(\varepsilon_2^k-\varepsilon_5^k)}-\frac {t_k\delta_{\text{M}}^2}{||\bm g_k||^2}\right]^+,  \forall k,
\end{equation}
where  $\varepsilon_2$, $\varepsilon_3^k$, $\varepsilon_5^k$, and $\varepsilon_6^k$ represent non-negative Lagrange multipliers associated with the corresponding constraints of problem (\ref{p33a}).

 \textit{Proof:} Please see Appendix F.

\textbf{\textit{Remark 5:}} In the max-min fairness regime, the optimization prioritizes the weakest WS by maximizing the minimum achievable utility across all WSs.  As indicated in (\ref{la3}), this ensures that every WS achieves at least a guaranteed service level, completely eliminating the resource disparity observed in Case I.  Compared to the other regimes, (\ref{ot2}) and (\ref{of2}) show that
 $t_k^*$ and $f_k^*$ are strongly influenced by the fairness multiplier $\lambda_6^k$, with more resources allocated to disadvantaged WSs, even at the expense of those with stronger channels.
Furthermore,  (\ref{op2}) reveals that $\bar p_k$ is strategically increased for weak-channel users, guided by $\varepsilon_6^k$, to improve their offloading efficiency under poor link scenarios. Although this strategy guarantees equitable task completion across all WSs, it inevitably sacrifices overall system throughput. Therefore, max-min fairness is particularly well-suited to mission-critical IoT applications, such as industrial automation, emergency networks, or public services, where ensuring the worst-case performance is a higher priority than maximizing total system efficiency.

\begin{algorithm}[t]
	\caption{Max-min Fairness-Based Algorithm (MFBA)}
	\SetAlgoLined
	\KwIn{System parameters: $P_{\max}$, $T$, $\epsilon$, $f_k^{\max}$, $B$, $\delta_{\text{M}}^2$, etc.}
	\KwOut{Optimal solution $\{\bar P_k^*, t_k^*, \bar p_k^*, f_k^*, \gamma^*\}$.}
	
	Initialize $\{\bar P_k, t_k, \bar p_k, f_k, \gamma\}$ with feasible values;\\
	
	\Repeat{convergence of $\{\bar P_k, t_k, \bar p_k, f_k, \gamma\}$}{
		\textbf{Step 1:} Optimize $t_k$ and $f_k$ with fixed $\bar P_k$ and $\bar p_k$\\
		\quad Solve (\ref{ot2}) and (\ref{of2}) using the current Lagrange multipliers $\{\lambda_1^k, \lambda_2, \lambda_3^k, \lambda_4^k, \lambda_5^k, \lambda_6^k\}$.\\
		
		\textbf{Step 2:} Optimize $\bar p_k$ and $\bar P_k$ with fixed $t_k$ and $f_k$\\
		\quad Solve (\ref{op2}) and (\ref{obp}) using updated multipliers $\{\varepsilon_1^k, \varepsilon_2^k, \varepsilon_3^k, \varepsilon_4^k, \varepsilon_5^k, \varepsilon_6^k\}$.\\
		
		\textbf{Step 3:} Update slack variable $\gamma$\\
		\quad Set $\gamma = \min_{\forall k} \bar R_k(\bar P_k, t_k, \bar p_k, f_k)$ according to (\ref{la3}).
	}
	
	\textbf{Step 4:} Recover original variables according to (\ref{os}).\\
	\KwResult{Return $\{P_k^*, t_k^*, p_k^*, f_k^*\}$.}
\end{algorithm}

The max-min fairness-based algorithm (MFBA) is summarized in \textbf{Algorithm 3}, which is similar to \textbf{Algorithm 1}. The main difference is that we need to upload the slack variable $\gamma$ according to (\ref{la3}) in each iteration until it converges.

\subsection{Offloading capacity enhancement with CER}

To analytically unveil the role of CER in enhancing offloading capacity, we consider a simplified case where each WS adopts full offloading, i.e., all harvested energy is devoted to data transmission. To isolate the CER effect among WSs, we assume equal time division and constant PS transmit power, namely $t_k = 1/K$ and $P_k = P_0$ for all $k$. Under these settings, the offloaded data sizes with and without CER are respectively given by
\begin{equation} \label{sc1}
	R_k^{\text{w. ER}}= \frac{1}{K} \log_2 \left(1+ \dfrac{K||\bm g_k||^2 E_k^{\text{w. ER}}}{\delta_{\text{M}}^2} \right), 
\end{equation} 
\begin{equation} \label{sc2}
	R_k^{\text{w.o. ER}}= \frac{1}{K} \log_2 \left(1+ \dfrac{K||\bm g_k||^2 E_k^{\text{w.o. ER}}}{\delta_{\text{M}}^2} \right), 
\end{equation} 
where 
\begin{equation}
	E_k^{\text{w. ER}}= \sum\limits_{i=1,1\neq k}^K \frac{1}{K} \eta   P_0 |h_k|^2+\sum\limits_{i=1,1\neq k}^K \frac{1}{K} \eta  p_i |g_{i,k}|^2,
\end{equation}
and 
\begin{equation}
	E_k^{\text{w.o. ER}}=\sum\limits_{i=1,1\neq k}^K  \frac{1}{K}  \eta P_0 |h_k|^2,
\end{equation}
represent the total harvested energy with and without CER, respectively.

The improvement in offloading capacity due to CER can thus be quantified as 
\begin{equation}  \label{sc5}
	\begin{aligned}
		R_k^{\text{Gap}}& =R_k^{\text{w. ER}}-R_k^{\text{w.o. ER}}\\
		&=\frac{1}{K} \log_2\left(\frac{\delta_{\text{M}}^2+K||\bm g_k||^2E_k^{\text{w. ER}}}{\delta_{\text{M}}^2+K||\bm g_k||^2E_k^{\text{w.o. ER}}}\right)\\
		&  \overset{(a)}{\approx} \frac{1}{K}  \log_2\left( \frac{E_k^{\text{w. ER}}}{E_k^{\text{w.o. ER}}} \right)\\
		&= \frac{1}{K} \log_2\left( 1+ \frac{ \sum\limits_{i=1,1\neq k}^K  p_i |g_{i,k}|^2}{\sum\limits_{i=1,1\neq k}^K  P_0 |h_k|^2} \right),\\		
	\end{aligned} 
\end{equation}
where approximation ($a$) assumes $\delta_{\text{M}}^2 \approx 0$, given that the noise power is negligible compared to the transmit signal. This approximation simplifies further derivation.
 
 \textbf{\textit{Remark 6}}: From (\ref{sc5}), it is clear that $R_k^{\text{gap}}$ is strongly influenced by the inter-WS channel gain $g_{i,k}$. When $g_{i,k}$ is small, the amount of energy that can be recycled is limited, leading to a smaller performance gap between CER-enabled and non-CER schemes. This implies that CER yields greater benefits in scenarios where WSs are spatially concentrated and inter-WS links are strong. Furthermore, $R_k^{\text{gap}}$ increases with the WS transmit power $p_i$, as higher transmit power enhances EH opportunities among WSs. The advantage of CER is also amplified when the direct PS signal strength, $P_0 |h_k|^2$, is weak, underscoring the importance of energy cooperation in improving offloading throughput under energy-constrained conditions. These analytical results indicate that CER is particularly effective in dense WS deployments or in networks with limited PS transmit power, a conclusion further corroborated by the simulation results in the next section.

\section{Numerical Results}

In this section, we present numerical results to evaluate the performance of the proposed algorithms. The considered WPCN-assisted MEC system consists of one PS, one AP equipped with 4 receive antennas, and four single-antenna WSs. The distances between the PS and each WS, as well as between each WS and the AP, are within 15 m, while the inter-WS distance is within 5 m. Large-scale fading is modeled by distance-dependent path loss, and small-scale fading follows a Rayleigh distribution for all channels. Specifically, the channel between the PS and the $k$-th WS is modeled as $h_k=\rho d_k^{-\beta}$, where $d_k$ denotes the transmission distance, $\beta=2.2$ is the path loss exponent, and $\rho$ is a Rayleigh random variable with unit variance \cite{10486847}. For each WS, the EH efficiency is assumed to be $\eta=0.8$ \cite{9888066}. The frame duration is normalized as $T=1$ s.  Unless otherwise stated, the remaining system parameters follow \cite{9312671,10301686,9919310}: $C_k=1000$ cycles/bit, $\phi_k=10^{-30}$,  $\delta_{\text{M}}^2=-90$ dBm, $B=1$ KHz, $P_{\max}=1$ W, $f_k^{\max}=1$ MHz, $R_k^{\min}=100$ bits.

To comprehensively evaluate the effectiveness of the proposed algorithms, we introduce the following benchmark schemes for performance comparison:
\begin{itemize}
    \item Full local-computing algorithm (FLCA) \cite{9312671}: In this algorithm, all WSs execute their computation tasks entirely using local processing resources, without any task offloading to the MEC server. 
    \item Full computation-offloading algorithm (FCOA) \cite{10444003}: In this algorithm, all computation tasks generated by the WSs are completely offloaded to the MEC server, and no local computation is performed. 
    \item Non-cooperative ER algorithm (NERA) \cite{9887822}: In this algorithm, the WSs harvest energy only from the dedicated PS, while CER among WSs is disabled. 
\end{itemize}

\subsection{The convergence of the proposed algorithms}

\begin{figure}[t]
	\centerline{\includegraphics[width=3in]{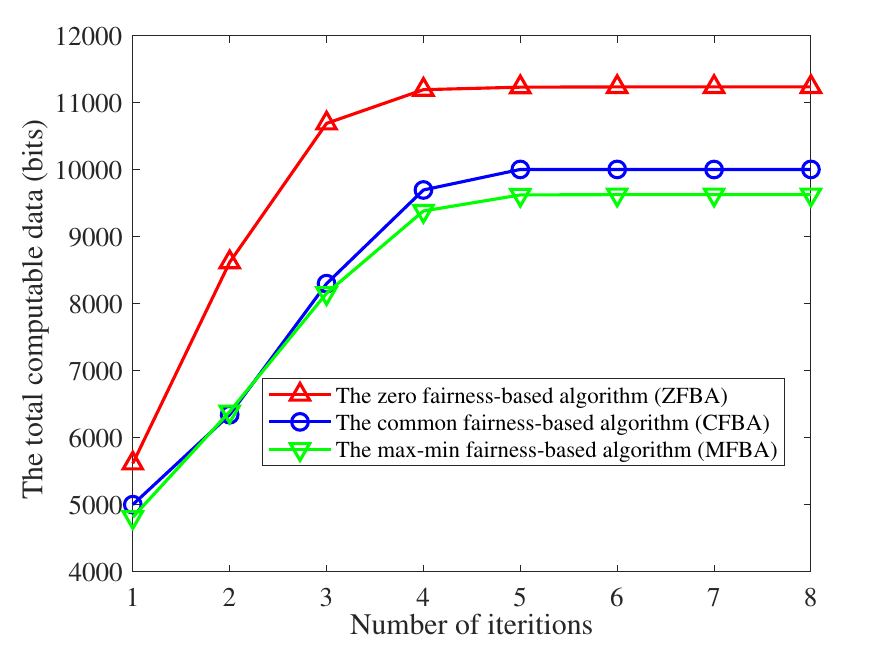}}
	\caption{The convergence of the proposed algorithms.}
	\label{fig2}
\end{figure} 

Fig. \ref{fig2} illustrates the convergence behavior of the proposed algorithms. As observed, all three algorithms exhibit rapid convergence in 4 to 5 iterations, confirming their computational efficiency. In particular, the ZFBA achieves the highest total computable data, as it allocates resources without fairness constraints. In contrast, other algorithms deliberately trade off part of the overall performance to ensure a more equitable distribution of resources among WSs, which demonstrates a classical balance between system throughput and user fairness in resource allocation.

\subsection{The fairness of the proposed algorithms}

\begin{figure}[t]
	\centerline{\includegraphics[width=3in]{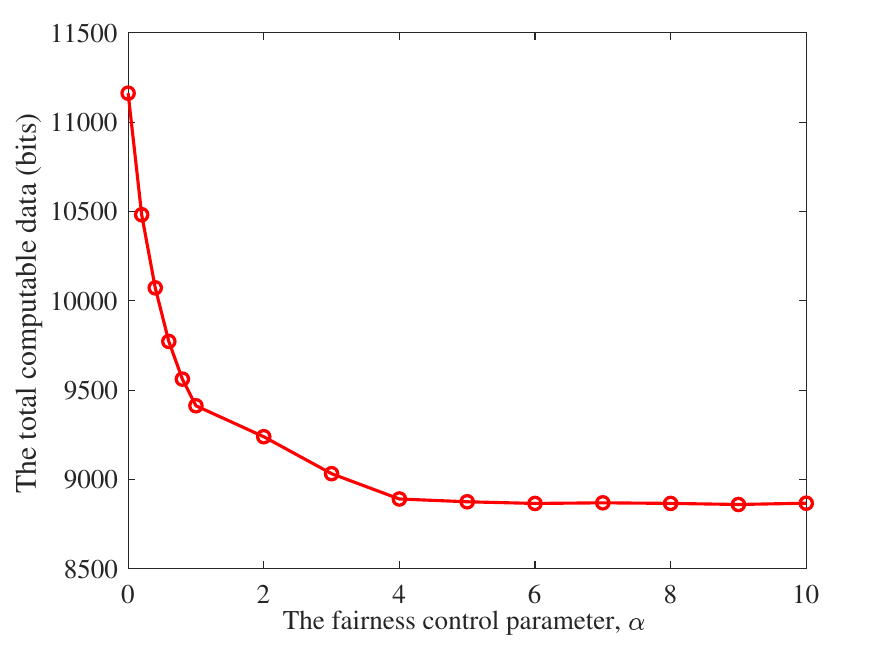}}
	\caption{The total computable data versus the fairness control parameter $\alpha$.}
	\label{fig3}
\end{figure} 

\begin{figure}[t]
	\centerline{\includegraphics[width=3in]{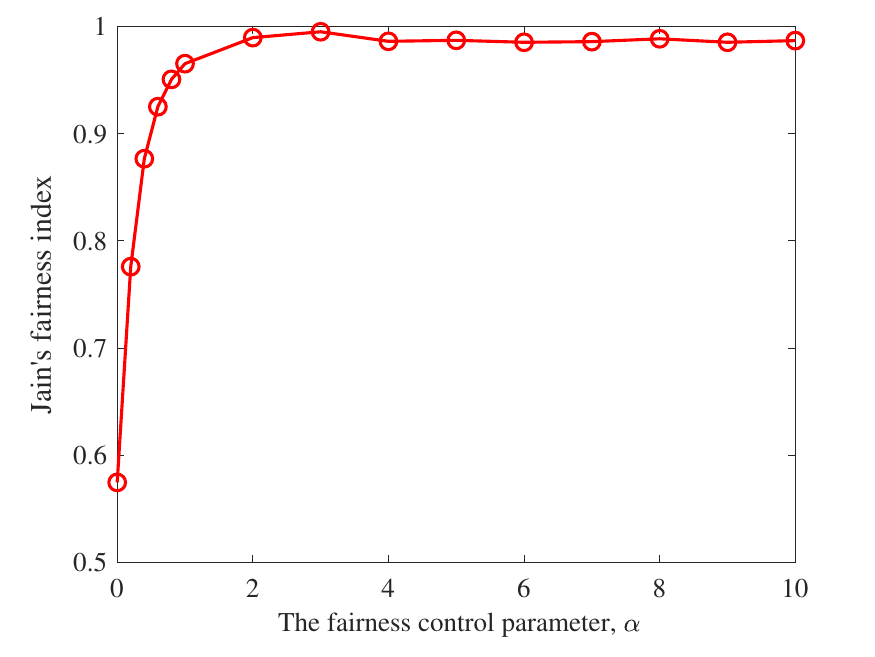}}
	\caption{The Jain's fairness index versus the fairness control parameter $\alpha$.}
	\label{fig4}
\end{figure} 

Fig. \ref{fig3} and Fig. \ref{fig4} jointly demonstrate the tradeoff between system efficiency and user fairness under varying fairness control parameter $\alpha$.  Specifically, Fig. \ref{fig3} shows that increasing $\alpha$ from 0 to 10 results in a steady decline in the total computable data, with performance stabilizing when $\alpha \geq 5$. This reflects a shift from an efficiency-oriented allocation strategy to a fairness-aware one, where resource concentration is increasingly regulated. Complementarily, Fig. \ref{fig4} presents the corresponding Jain’s fairness index, defined as ${\text{JFI}}=(\sum_{k=1}^K R_k)^2/(K\sum_{k=1}^K R_k^2)$ \cite{9625847}, which improves rapidly with increasing $\alpha$, particularly in the low-$\alpha$ region (i.e., $\alpha < 2$). The index saturates around 0.97–0.98 beyond $\alpha \approx 2$, indicating that near-optimal fairness can be achieved with moderate enforcement. However, this comes at the cost of performance degradation, as shown in Fig. \ref{fig3}.

These trends collectively highlight a diminishing-return phenomenon: while small values of $\alpha$ lead to substantial fairness gains with minimal performance loss, further increasing $\alpha$ yields marginal fairness improvement but more pronounced efficiency degradation. Therefore, choosing $\alpha$ in a moderate range (e.g., 2–5) can effectively balance fairness and performance, ensuring both equitable user treatment and acceptable system throughput.

\begin{figure}[t]
	\centerline{\includegraphics[width=3in]{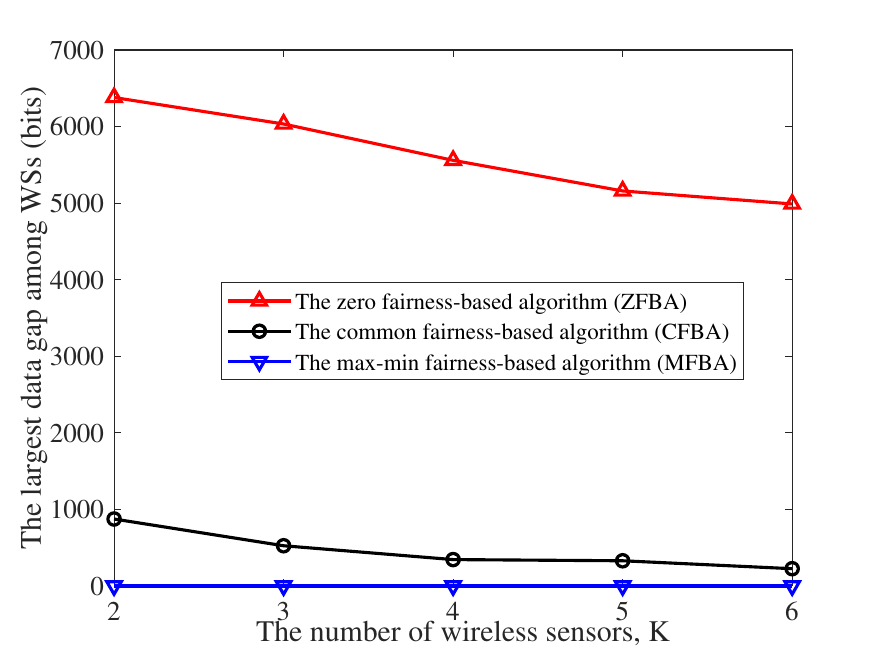}}
	\caption{The largest data gap versus the number of WSs.}
	\label{fig5}
\end{figure} 

 Fig. \ref{fig5} illustrates the largest data gap among WSs for the proposed algorithms under the number of WSs ($K$). As expected, the ZFBA exhibits a consistently large gap, exceeding 6000 bits when only two WSs are involved. This is because the ZFBA focuses solely on maximizing total data size, often allocating excessive resources to a few high-performing users while neglecting others, especially in small-scale systems.
In contrast, the MFBA maintains an almost negligible gap in data size, demonstrating strong fairness enforcement regardless of network size. The CFBA strikes a middle ground, offering a significant reduction in the data gap compared to the ZFBA while preserving reasonable throughput efficiency. Moreover, as $K$ increases, the data gap under all three algorithms tends to decrease. This is attributed to the increased diversity in user conditions and the finer granularity in resource distribution, which inherently mitigates extreme allocations. Nevertheless, the differences between the algorithms remain significant, highlighting the importance of fairness-aware designs in systems with limited user populations, where imbalance is most pronounced.

\subsection{The performance of the proposed algorithms}

\subsubsection{The scheme comparison with the proposed schemes}

\begin{figure}[t]
	\centerline{\includegraphics[width=3in]{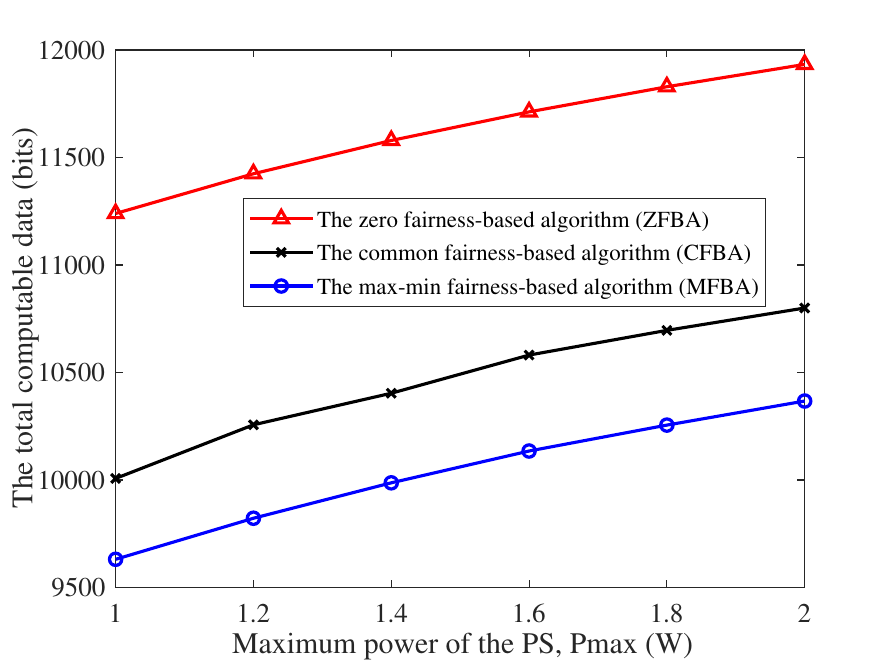}}
	\caption{The total computable data versus the maximum power of the PS.}
	\label{fig6}
\end{figure} 

\begin{figure}[t]
	\centerline{\includegraphics[width=3in]{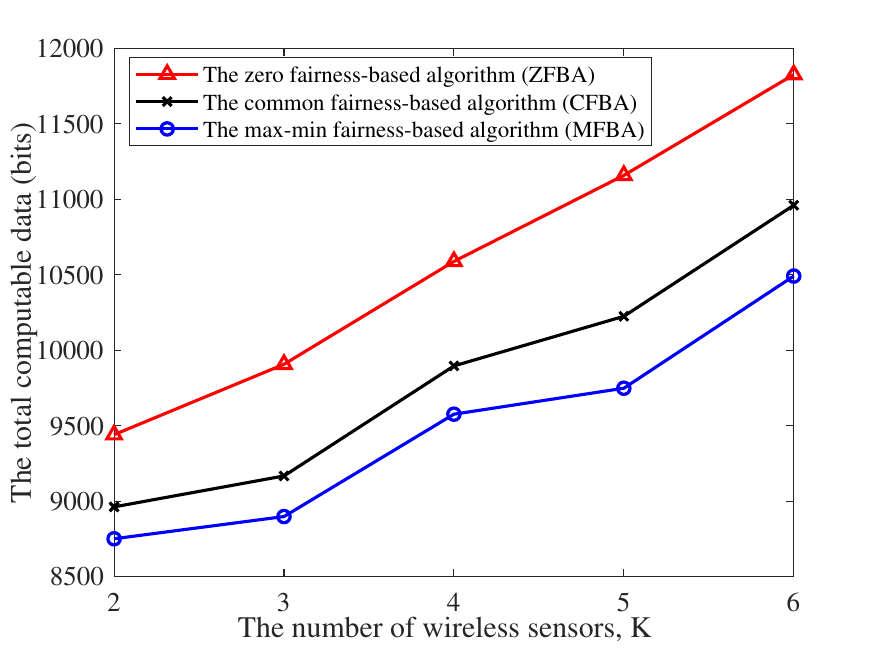}}
	\caption{The total computable data versus the number of WSs.}
	\label{fig7}
\end{figure}

\begin{figure}[t]
	\centerline{\includegraphics[width=3in]{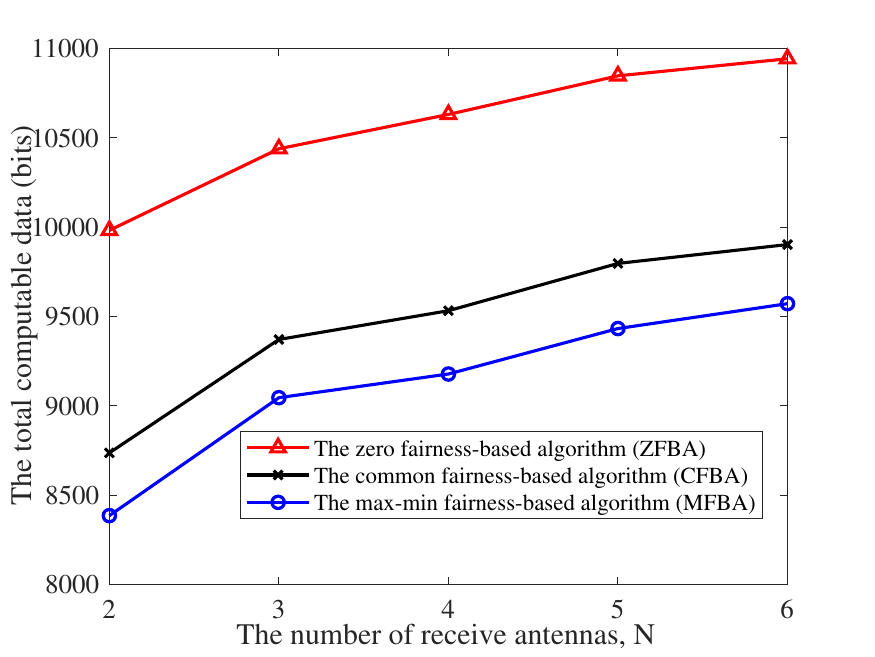}}
	\caption{The total computable data versus the number of receive antennas.}
	\label{fig8}
\end{figure}

 Figs. \ref{fig6}–\ref{fig8} jointly investigate the performance dynamics of the three proposed algorithms, i.e., ZFBA, CFBA, and MFBA, under varying system parameters, namely the maximum power budget ($P_{\max}$), the number of WSs ($K$), and the number of receive antennas ($N$).

As shown in Fig. \ref{fig6}, increasing $P_{\max}$ yields a nearly linear rise in total computable data across all algorithms. This trend reflects the direct benefit of enhanced energy availability on both local computation and wireless transmission capabilities. Notably, the ZFBA exhibits the steepest growth, aggressively utilizing the extra energy to boost throughput by favoring high-efficiency users. In contrast, fairness-based algorithms show more moderate gains due to their regulated resource allocation, particularly the MFBA, which prioritizes uniform performance distribution. This result highlights that energy acts as a performance bottleneck, and releasing this constraint enables efficiency-centric algorithms to fully realize their potential, while fairness constraints inherently limit such exploitation.

In Fig. \ref{fig7}, scaling $K$ also significantly improves total computable data for all schemes. The improvement stems from greater computational diversity and enhanced scheduling flexibility. The performance boost is especially pronounced for the ZFBA, which selectively allocates tasks to strong nodes, resulting in larger gaps between the ZFBA and fairness-aware methods. This disparity implies that user scaling, although beneficial for overall system capacity, may exacerbate inequities if fairness is not adequately enforced. Interestingly, the marginal gains for the CFBA and the MFBA grow more rapidly after $K$ exceeds 3, suggesting that fairness-aware scheduling benefits from richer user pools with CER, where task redistribution becomes more efficient and less penalizing.

Fig. \ref{fig8} presents the effect of increasing the number of receive antennas, which enhances signal reception quality. All algorithms benefit steadily from this infrastructure-side enhancement. Importantly, the performance gaps among the three algorithms remain relatively consistent across the antenna configurations, indicating that improving system-side capabilities (e.g., via more antennas) yields throughput gains without significantly intensifying resource inequality. This insight reveals a valuable design direction: infrastructure expansion is more fairness-compatible than simply increasing energy budgets, making it a more sustainable strategy for enhancing performance under fairness constraints.

\subsubsection{The scheme comparison with the other benchmark schemes}

\begin{figure}[t]
	\centerline{\includegraphics[width=3in]{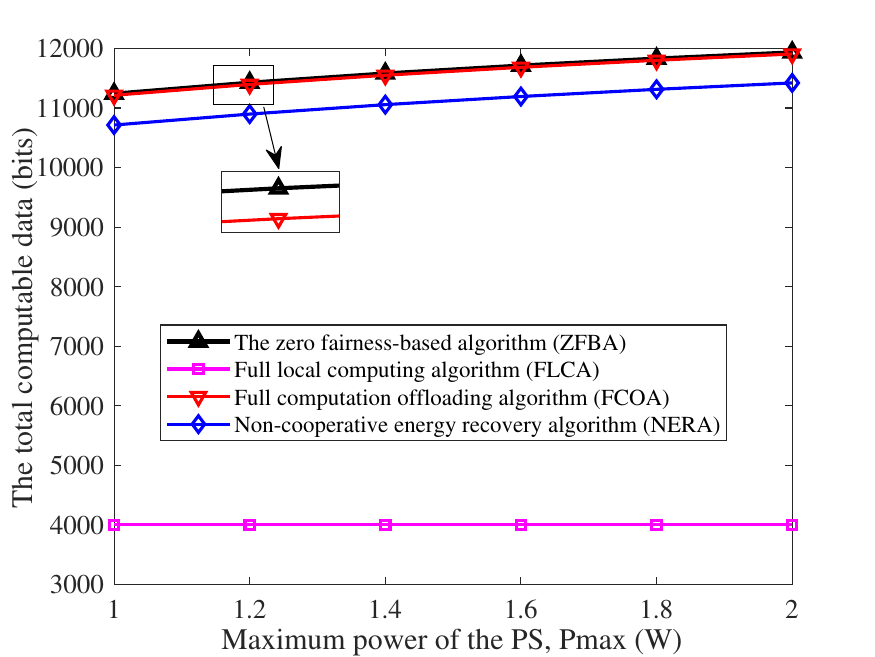}}
	\caption{The total computable data versus the maximum power of the PS.}
	\label{fig9}
\end{figure}

\begin{figure}[t]
	\centerline{\includegraphics[width=3in]{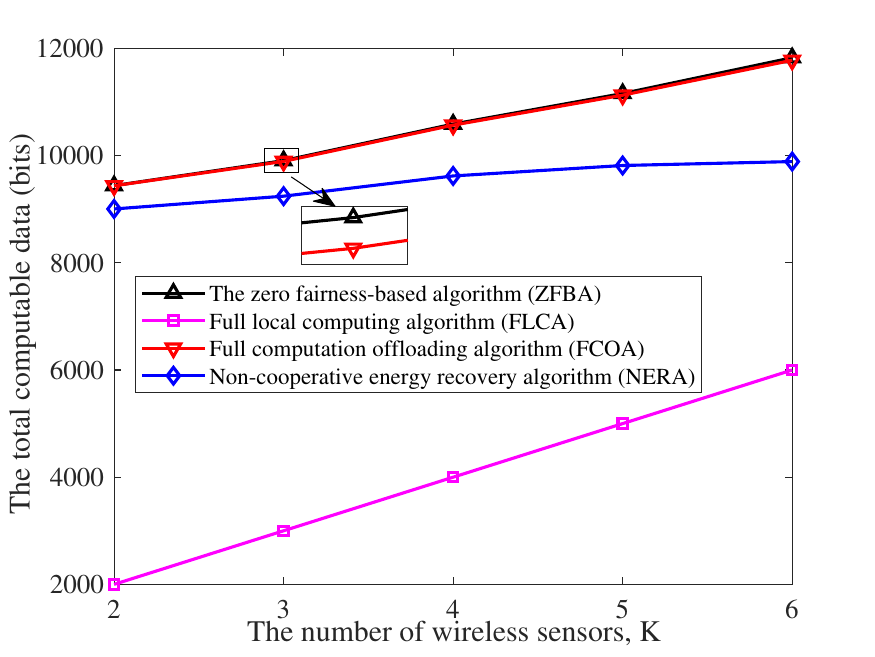}}
	\caption{The total computable data versus the number of WSs.}
	\label{fig10}
\end{figure}

\begin{figure}[t]
	\centerline{\includegraphics[width=3in]{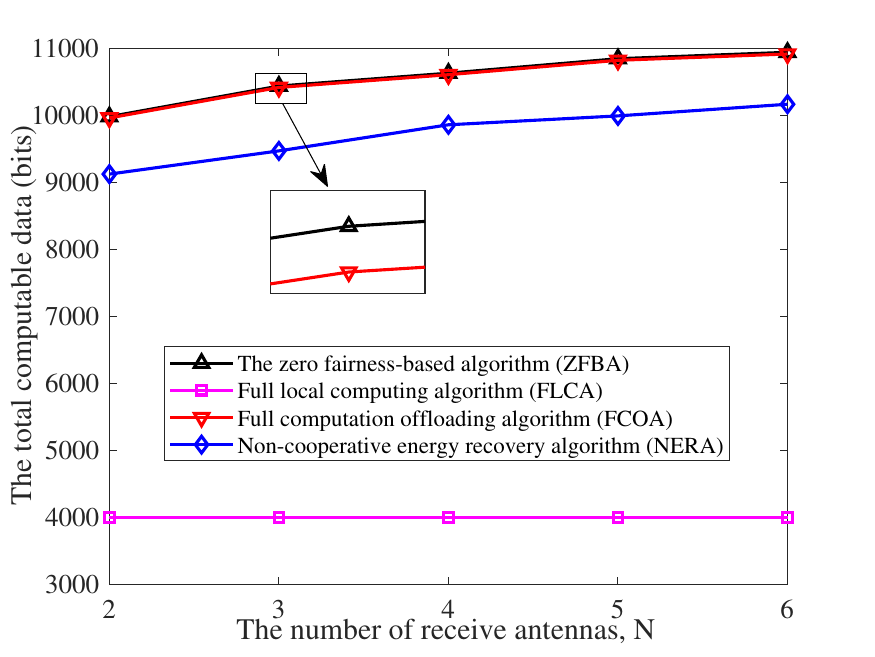}}
	\caption{The total computable data versus the number of receive antennas.}
	\label{fig11}
\end{figure}

To further validate the effectiveness of the proposed algorithms, we conduct a comparative study against several representative benchmark schemes, including FLCA, FCOA, and NERA, in Figs. \ref{fig9}–\ref{fig11}.  To ensure a fair and consistent comparison, all schemes are assessed under the same fairness-neutral condition, specifically by setting $\alpha=0$. This eliminates the influence of fairness prioritization, allowing a direct assessment of each algorithm’s efficiency in terms of total computable data and energy utilization. 

As shown in Fig. \ref{fig9}, increasing $P_{\max}$ steadily enhances the total computable data for all algorithms, except the FLCA, which remains nearly flat due to the lack of offloading and the limited computation capacity. The ZFBA consistently achieves the best performance, taking full advantage of energy increments via joint EH, task scheduling, and offloading optimization. Compared to the FCOA and the NERA, the ZFBA exhibits a noticeable performance gain by flexibly balancing local computing and offloading while enabling cooperative energy recovery among WSs. This validates the design advantage of integrating CER with adaptive task allocation. Meanwhile, the performance of the FCOA slightly lags due to its rigid offloading behavior, which overlooks local computation potential and may suffer under limited MEC resources. The NERA, though capable of offloading, is restricted by the lack of energy cooperation, showing moderate performance improvements.

Moving to Fig. \ref{fig10}, we observe that expanding $K$ significantly boosts system capacity, particularly for the ZFBA and the FCOA. The ZFBA achieves the most substantial improvements by selectively leveraging new nodes with strong computing and EH capabilities. The result illustrates that a richer user pool offers more granular scheduling flexibility and better energy-task matching opportunities. The NERA, although benefiting from increased nodes, is limited by its isolated energy supply design. Meanwhile, the FLCA shows linear but limited scaling due to its fixed local-only execution strategy. Overall, it can be seen that multi-user diversity greatly benefits offloading-capable and cooperative designs, reinforcing the role of resource coordination in scalability.

In Fig. \ref{fig11}, the impact of increasing $N$ is analyzed. As expected, all offloading-enabled algorithms benefit from enhanced signal reception, which facilitates higher-quality task transmission. The ZFBA maintains its superiority across the board, consistently outperforming the alternatives by optimizing both channel conditions and resource distribution. The FCOA also gains steadily but lacks adaptive balancing between local and remote execution. The improvement of the NERA is less pronounced, primarily due to its restrictive energy sourcing model. Once again, the FLCA shows negligible response to communication-side upgrades, highlighting its inherent limitation in offloading-dominant environments. Interestingly, the relative performance gaps across the offloading-enabled algorithms remain stable, implying that infrastructure enhancements, i.e., antenna upgrades, tend to benefit all schemes fairly without intensifying disparities.

\section{Conclusion}

This paper studied a CER architecture for wireless-powered MEC networks under an $\alpha$-fairness-driven optimization framework. The proposed model jointly accounts for local computing and task offloading, while embedding fairness control into the objective of maximizing computable data. By reformulating the original non-convex problem into a tractable structure, closed-form solutions were derived for three representative fairness regimes via dual decomposition and alternating optimization. Simulation results validated the theoretical analysis and showed that, compared to conventional baselines, the proposed framework achieves significant improvements in total computable data and fairness, while offering superior adaptability and scalability across diverse MEC scenarios.

	\begin{appendices}

\renewcommand{\theequation}{\Alph{section}\arabic{equation}}
	
	\section{Proof of  Theorem 1}

According to the property of perspective
function, $f(x,y)\triangleq x\log(1+y/x)$ is concave with respect to $x$ and $y$. Thus, $t_k\log_2( 1+{\bar p_k||\bm g_k||^2}/{(t_k\delta_{\text{M}}^2)})$ is concave with respect to $t_k$ and $\bar p_k$. Besides, 	$R_k^{\text{LC}}={T f_k}/{C_k}$ is a linear function related to $f_k$. Thus, $\bar R_k$ is concave with respect to $t_k$, $\bar p_k$, and $f_k$.

According to (\ref{f1}), we have the following derivations, i.e., 
\setcounter{equation}{0}
\begin{equation} \label{ap11}
	u'_\alpha(x)=x^{-\alpha} \ge 0, \forall \alpha \ge 0,x \ge 0,\\
\end{equation}
\begin{equation} \label{ap12}
	u''_\alpha(x)=-x^{-\alpha-1} \le 0, \forall \alpha \ge 0,x \ge 0,\\
\end{equation}
which demonstrates that $u_\alpha(x)$ is concave and nondecreasing. Then, based on the property of scalar composition, $u_\alpha(\bar R_k)$ is also concave. Thus, the objective function of problem (\ref{p3}) is concave with respect to $\bar P_k$,  $t_k$, $\bar p_k$, and $f_k$. 

The proof is complete.

\section{Proof of  Lemma 1}

 Since $t_k$ is only involved in ${u_\alpha }(\bar R_k)$ in the objective function of problem (\ref{p3}), here we first discuss the first-order derivative of  $\bar R_k$ with the respect to $t_k$, which can be expressed as 
 \textbf{\setcounter{equation}{0}
 \renewcommand{\theequation}{B.\arabic{equation}}}
\begin{equation} \label{ap21}
	\begin{aligned}
			\dfrac{\partial \bar R_k}{\partial t_k}=&\frac{B}{\ln 2} \ln \left( 1+\dfrac{\bar p_k||\bm g_k||^2}{t_k\delta_{\text{M}}^2}\right)\\
			&-\frac{B}{\ln 2} \dfrac{\bar p_k||\bm g_k||^2}{\bar p_k||\bm g_k||^2+t_k\delta_{\text{M}}^2}.\\
	\end{aligned}
\end{equation}

By defining function 
\begin{equation} \label{ap22}
	f_t(x)= \frac{1}{\ln 2} \left( \ln (1+x)-\frac{x}{1+x}\right), \forall x\ge 0,
\end{equation}
we can obtain 
\begin{equation} \label{ap23}
	f'_t(x)=\frac{1}{\ln 2}\dfrac{x}{(x+1)^2} \ge 0, \forall x >0.\\
\end{equation}

Therefore, $f_t(x)$ is a monotonically increasing function, satisfying $f_t(x)>f_t(0)=0$. According to (\ref{ap21}) and (\ref{ap22}), we have $\dfrac{\partial \bar R_k}{\partial t_k}=Bf_t(\dfrac{\bar p_k||\bm g_k||^2}{t_k\delta_{\text{M}}^2})$. Then, we have $\dfrac{\partial \bar R_k}{\partial t_k}\ge 0$ and $\bar R_k$ is increasing with $t_k$. Besides, based on (\ref{ap11}), ${u_\alpha }(\bar R_k)$ is also increasing with $t_k$.
Thus,  the objective function of problem (\ref{p3}) is always improved with the increasing $t_k^*$. That is to say,  $\sum_{k=1}^K t_k^* < T-\epsilon$ can not hold. 

The proof is complete.

\section{Proof of  Theorem 2}

Given fixed transmit powers ${\bar P_k}$ and ${\bar p_k}$, the Lagrangian function of problem (\ref{p31}) can be expressed as
\setcounter{equation}{0}
\begin{equation} \label{apc1}
	\begin{aligned}
		&\mathcal L_1 (t_k, f_k, \omega_1^k, \omega_2, \omega_3^k, \omega_4^k, \omega_5^k)\\
		&= \sum\limits_{k=1}^K \bar R_k +  \sum\limits_{k=1}^K\omega_1^k(P_{\max}t_k{-}\bar P_k)\\
		&
		+\omega_2 (T-\epsilon-\sum_{k=1}^K t_k ) + \sum\limits_{k=1}^K\omega_3^k(\bar E_k^{\text{EH}}-\bar E_k^{\text{EC}})\\
		&+\sum\limits_{k=1}^K\omega_4^k(f_k^{\max}-f_k) + \sum\limits_{k=1}^K\omega_5^k (\bar R_k- R_k^{\min}),\\
	\end{aligned}
\end{equation}
where $\omega_1^k$, $\omega_3^k$, $\omega_4^k$, and $\omega_5^k$ represent non-negative Lagrange multipliers associated with the corresponding constraints of problem (\ref{p31}). Besides, by Lemma 1, $C_2$ is tight at optimum, hence Lagrange multiplier $\omega_2 >0$.

By applying the Karush–Kuhn–Tucker (KKT) conditions, the first-order derivatives with respect to $t_k$ and $f_k$ are given by
\begin{equation} \label{apc2}
	\begin{aligned}
		\dfrac{\partial \mathcal L_1}{\partial t_k}&= (1+\omega_5^k)Bf_t\left( \dfrac{\bar p_k||\bm g_k||^2}{t_k\delta_{\text{M}}^2}\right) +\omega_1^kP_{\max}-\omega_2=0.
	\end{aligned}
\end{equation}
\begin{equation} \label{apc3}
	\dfrac{\partial \mathcal L_1}{\partial f_k}= (1+\omega_5^k)\frac{ T}{C_k}-3\omega_3^k T\phi_kf^2-\omega_4^k=0.
\end{equation}

From (\ref{apc2}), together with the complementary slackness conditions, the optimal transmission time $t_k^*$ can be obtained as (\ref{ot}), where $f_t^{-1}(y)$ is the inverse function of $f_t(x)$ defined as
\begin{equation} \label{apc4a}
	f_t^{-1}(y)=e^{W\left(-\dfrac{1}{e^{1+y\ln 2}}\right)+1+y\ln 2 }-1,
\end{equation}
where $W(x)$ is the Lambert-$W$ function \cite{lwf}.

Besides, from (\ref{apc3}), the optimal CPU frequency $f_k^*$ can be derived as in (\ref{of}).
 
Then, with $t_k$ and $f_k$ fixed, the Lagrangian function for optimizing ${\bar P_k, \bar p_k}$ can be expressed as
\begin{equation} \label{apc6}
	\begin{aligned}
		&\mathcal L_2 (\bar P_k, \bar p_k, \mu_1^k, \mu_2^k, \mu_3^k, \mu_4^k, \mu_5^k)\\
		&= \sum\limits_{k=1}^K  \bar R_k +\sum\limits_{k=1}^K\mu_1^k(P_{\max}t_k{-}\bar P_k) \\
		&+ \sum\limits_{k=1}^K\mu_2^k(\bar E_k^{\text{EH}}-\bar E_k^{\text{EC}}) + \sum\limits_{k=1}^K\mu_3^k (\bar R_k- R_k^{\min})\\
		&
		+\sum\limits_{k=1}^K\mu_4^k \bar P_k+\sum\limits_{k=1}^K\mu_5^k \bar p_k,
	\end{aligned}
\end{equation}
where $\mu_1^k$, $\mu_2^k$, $\mu_3^k$, $\mu_4^k$, and $\mu_5^k$  represent non-negative Lagrange multipliers associated with the corresponding constraints of problem (\ref{p31}). 

 By deducing the first-order derivative of $\bar p_k$ and $\bar P_k$, based on KKT condition, the optimal solutions of $\bar p_k$ and $\bar P_k$ should satisfy
\begin{equation} \label{apc7}
	\begin{aligned}
		\dfrac{\partial \mathcal L_2}{\partial \bar p_k}&=\frac{(1+\mu_3^k) Bt_k}{\ln 2}\frac{||\bm g_k||^2}{\bar p_k||\bm g_k||^2+t_k\delta_{\text{M}}^2}
		-\mu_2^k+\mu_5^k=0.
	\end{aligned}
\end{equation}
\begin{equation} \label{apc8} 
		\dfrac{\partial \mathcal L_2}{\partial \bar P_k}= -\mu_1^k+\mu_4^k=0.
\end{equation}

From (\ref{apc7}), the optimal $\bar p_k^*$ can be expressed as (\ref{op}) using complementary slackness. Moreover, from (\ref{apc8}), it follows that $\mu_4^k = \mu_1^k$, implying that $\bar P_k$ can be freely selected in the feasible region $[0, P_{\max}t_k]$. Since increasing $\bar P_k$ cannot deteriorate the objective, the optimal choice can be set as in (\ref{obp}). The proof is complete.

\section{Proof of Lemma 2}

Because $u_\alpha(\chi_k)$ is strictly increasing with respect to $\chi_k$, it is evident that the optimal solution of problem (\ref{p32a}) must satisfy the equality condition of $C_8$, i.e., $\chi_k = \bar R_k$ for all $k$. For completeness, we provide a contradiction-based proof as follows.

Assume, for contradiction, that $\bar R_k(\bar P_k^*, t_k^*, \bar p_k^*, f_k^*) \neq \chi_k^*$ for some $k$. 

\textit{Case I}: If $\bar R_k(\bar P_k^*, t_k^*, \bar p_k^*, f_k^*) > \chi_k^*$, then it is possible to increase $\chi_k$ up to $\bar R_k(\bar P_k^*, t_k^*, \bar p_k^*, f_k^*)$ without violating $C_8$, which increases the objective since $u_\alpha(\cdot)$ is monotonic. This contradicts the optimality of $\chi_k^*$.

\textit{Case II}: If $\bar R_k(\bar P_k^*, t_k^*, \bar p_k^*, f_k^*) < \chi_k^*$, then constraint $C_8$ is violated.

Therefore, in both cases a contradiction arises, which implies that $\bar R_k(\bar P_k^*, t_k^*, \bar p_k^*, f_k^*) = \chi_k^*$ for all $k$. Consequently, problem (\ref{p32a}) is equivalent to problem (\ref{p3}). The proof is complete.

\section{Proof of  Theorem 3}
 With the given $\bar P_k$ and $\bar p_k$, the Lagrange function of problem (\ref{p32a}) can be written by
 \setcounter{equation}{0}
\begin{equation} \label{apd1}
	\begin{aligned}
		&\mathcal L_3 (\chi_k, t_k, f_k, \zeta_1^k, \zeta_2, \zeta_3^k, \zeta_4^k, \zeta_5^k, \zeta_6^k)\\
		&= \sum\limits_{k=1}^K {u_\alpha }(\chi_k)+  \sum\limits_{k=1}^K\zeta_1^k(P_{\max}t_k{-}\bar P_k)
		 \\
		&+\zeta_2 (T-\epsilon-\sum_{k=1}^K t_k ) + \sum\limits_{k=1}^K\zeta_3^k(\bar E_k^{\text{EH}}-\bar E_k^{\text{EC}})\\
		&+\sum\limits_{k=1}^K\zeta_4^k(f_k^{\max}-f_k)+ \sum\limits_{k=1}^K\zeta_5^k (\chi_k- R_k^{\min})\\
		&+\sum\limits_{k=1}^K\zeta_6^k(\bar R_k-\chi_k),
	\end{aligned}
\end{equation}
where $\zeta_1^k$, $\zeta_2$, $\zeta_3^k$, $\zeta_4^k$, $\zeta_5^k$, and $\zeta_6^k$ represent non-negative Lagrange multipliers associated with the corresponding constraints of problem (\ref{p32a}).

 Applying the KKT conditions, we obtain
  \begin{equation} \label{apd2}
 	\begin{aligned}
 		\dfrac{\partial \mathcal L_3}{\partial \chi_k}&= \chi_k^{-\alpha}+\zeta_5^k-\zeta_6^k=0.
 	\end{aligned} 
 \end{equation}
 \begin{equation} \label{apd3}
 	\begin{aligned}
 			\dfrac{\partial \mathcal L_3}{\partial t_k}&= \zeta_1^kP_{\max}-\zeta_2+\zeta_6^kBf_t\left( \dfrac{\bar p_k||\bm g_k||^2}{t_k\delta_{\text{M}}^2}\right)=0.
 	\end{aligned} 
 \end{equation}
  \begin{equation} \label{apd4}
 	\begin{aligned}
 		\dfrac{\partial \mathcal L_3}{\partial f_k}&= \zeta_6^k\frac{ T}{C_k}-3\zeta_3^kT\phi_kf^2-\zeta_4^k=0.
 	\end{aligned} 
 	\end{equation}

From (\ref{apd2})–(\ref{apd4}) with complementary slackness, the optimal $\chi_k$ is
\begin{equation} \label{apd2a}
	\chi_k= \left( {\zeta_6^k-\zeta_5^k}\right)^{-\dfrac{1}{\alpha}}, \forall k,
\end{equation}
and the closed-form solutions for $t_k^*$ and $f_k^*$ follow as in (\ref{ot1})–(\ref{of1}).

Then, with the given $t_k$, $f_k$,  the Lagrange function of problem (\ref{p32a}) can be written by
\begin{equation} \label{apd7}
	\begin{aligned}
		&\mathcal L_4 (\chi_k, \bar P_k, \bar p_k, \theta_1^k, \theta_2^k, \theta_3^k, \theta_4^k, \theta_5^k, \theta_6^k)\\
		&= \sum\limits_{k=1}^K   {u_\alpha }(\chi_k) +\sum\limits_{k=1}^K\theta_1^k(P_{\max}t_k{-}\bar P_k) \\
		&+ \sum\limits_{k=1}^K\theta_2^k(\bar E_k^{\text{EH}}-\bar E_k^{\text{EC}}) + \sum\limits_{k=1}^K\theta_3^k (\chi_k- R_k^{\min}) \\
		&
		+\sum\limits_{k=1}^K\theta_4^k(\bar R_k-\chi_k) +\sum\limits_{k=1}^K\theta_5^k \bar P_k+\sum\limits_{k=1}^K\theta_6^k \bar p_k,\\
	\end{aligned}
\end{equation}
where $\theta_1^k$, $\theta_2^k$, $\theta_3^k$, $\theta_4^k$, $\theta_5^k$, and $\theta_6^k$  represent non-negative Lagrange multipliers associated with the corresponding constraints of problem (\ref{p32a}). 

Taking derivatives and applying KKT conditions, we have
\begin{equation} \label{apd7a}
	\dfrac{\partial \mathcal L_4}{\partial \chi_k}= \chi_k^{-\alpha}+\theta_3^k-\theta_4^k=0,
\end{equation}
\begin{equation} \label{apd8}
	\dfrac{\partial \mathcal L_4}{\partial \bar p_k}= -\theta_2^k+\theta_4^k \frac{Bt_k}{\ln 2}\frac{||\bm g_k||^2}{\bar p_k||\bm g_k||^2+t_k\delta_{\text{M}}^2}+\theta_6^k=0.
\end{equation}
\begin{equation} \label{apd9} 
	\dfrac{\partial \mathcal L_4}{\partial \bar P_k}= -\theta_1^k+\theta_5^k=0.
\end{equation}

According to (\ref{apd7a})-(\ref{apd9}) with complementary slackness, the updated variables are
\begin{equation} \label{apd7b}
	\chi_k= \left( {\theta_4^k-\theta_3^k}\right)^{-\dfrac{1}{\alpha}}, \forall k,
\end{equation}
and the optimal $\bar p_k^*$ can be derived as (\ref{op1}). 
Moreover, from the derivative with respect to $\bar P_k$, it follows that $\theta_1^k=\theta_5^k$. Thus,  $\bar P_k^*$ takes the same optimal form as in (\ref{obp}). The proof is complete.

\section{Proof of  Theorem 4}
With the given $\bar P_k$ and $\bar p_k$, the Lagrange function of problem (\ref{p33a}) can be written by
\setcounter{equation}{0}
\begin{equation} \label{ape1}
	\begin{aligned}
		&\mathcal L_5 (\gamma, t_k, f_k, \lambda_1^k, \lambda_2, \lambda_3^k, \lambda_4^k,\lambda_5^k,\lambda_6^k)\\
		&= \gamma 
		+\sum\limits_{k=1}^K\lambda_1^k(P_{\max}t_k{-}\bar P_k)
		+\lambda_2 (T-\epsilon-\sum_{k=1}^K t_k )\\
		&+ \sum\limits_{k=1}^K\lambda_3^k(\bar E_k^{\text{EH}}-\bar E_k^{\text{EC}})+\sum\limits_{k=1}^K\lambda_4^k(f_k^{\max}-f_k)\\
		& + \sum\limits_{k=1}^K\lambda_5^k (\bar R_k- R_k^{\min})+\sum\limits_{k=1}^K\lambda_6^k ( \bar R_k- \gamma),\\
	\end{aligned}
\end{equation}
where $\lambda_1^k$, $\lambda_2$, $\lambda_3^k$, $\lambda_4^k$, $\lambda_5^k$, and $\lambda_6^k$ represent non-negative Lagrange multipliers associated with the corresponding constraints of problem (\ref{p33a}).

Taking the first-order derivatives and applying the KKT conditions, we have
\begin{equation} \label{ape2}
	\dfrac{\partial \mathcal L_5}{\partial \gamma}= 1-\sum\limits_{k=1}^K\lambda_6^k=0,
\end{equation}
\begin{equation} \label{ape3}
	\begin{aligned}
		\dfrac{\partial \mathcal L_5}{\partial t_k}&=  \lambda_1^kP_{\max}-\lambda_2+(\lambda_5^k+\lambda_6^k)Bf_t\left( \dfrac{\bar p_k||\bm g_k||^2}{t_k\delta_{\text{M}}^2}\right)=0,
	\end{aligned}
\end{equation}
\begin{equation} \label{ape4}
	\dfrac{\partial \mathcal L_5}{\partial f_k}= (\lambda_5^k+\lambda_6^k)\frac{ T}{C_k} -3\lambda_3^k T\phi_kf^2-\lambda_4^k=0.
\end{equation}

From (\ref{ape3}) and (\ref{ape4}), and applying complementary slackness, the optimal  $t_k^*$ and $f_k^*$ can be expressed as (\ref{ot2}) and (\ref{of2}), respectively.

Next, with fixed $t_k$ and $f_k$, the Lagrangian function for optimizing $\bar P_k$ and $\bar p_k$ becomes
\begin{equation} \label{ape8}
	\begin{aligned}
		&\mathcal L_6 (\gamma, \bar P_k, \bar p_k, \varepsilon_1^k, \varepsilon_2^k, \varepsilon_3^k, \varepsilon_4^k, \varepsilon_5^k, \varepsilon_6^k )\\
		&=\gamma
		+\sum\limits_{k=1}^K\varepsilon_1^k(P_{\max}t_k{-}\bar P_k)+ \sum\limits_{k=1}^K\varepsilon_2^k(\bar E_k^{\text{EH}}-\bar E_k^{\text{EC}})\\
		& + \sum\limits_{k=1}^K\varepsilon_3^k (\bar R_k- R_k^{\min})+\sum\limits_{k=1}^K\varepsilon_4^k \bar P_k+\sum\limits_{k=1}^K\varepsilon_5^k \bar p_k\\
		&+\sum\limits_{k=1}^K\varepsilon_6^k ( \bar R_k - \gamma),
	\end{aligned}
\end{equation}
where $\varepsilon_1^k$, $\varepsilon_2^k$, $\varepsilon_3^k$, $\varepsilon_4^k$, $\varepsilon_5^k$, and $\varepsilon_6^k$  represent non-negative Lagrange multipliers associated with the corresponding constraints of problem (\ref{p33a}). 

Taking derivatives and applying KKT conditions, we have
\begin{equation} \label{ape8a}
	\dfrac{\partial \mathcal L_6}{\partial \gamma}= 1-\sum\limits_{k=1}^K\varepsilon_6^k=0,
\end{equation}
\begin{equation} \label{ape9}
	\begin{aligned}
			\dfrac{\partial \mathcal L_6}{\partial \bar p_k}=& (\varepsilon_3^k+\varepsilon_6^k) \frac{ Bt_k}{\ln 2}\frac{||\bm g_k||^2}{\bar p_k||\bm g_k||^2+t_k\delta_{\text{M}}^2}{-}\varepsilon_6^k\psi\\
			& -\varepsilon_2^k+\varepsilon_5^k=0.
	\end{aligned}
\end{equation}
\begin{equation} \label{ape10} 
	\dfrac{\partial \mathcal L_6}{\partial \bar P_k}= \varepsilon_4^k-\varepsilon_1^k=0.
\end{equation}

Similarly, according to (\ref{ape9}) and (\ref{ape10}), the optimal $\bar p_k^*$ and $\bar P_k^*$ can be obtained as (\ref{op2}) and  (\ref{obp}), respectively. The proof is complete.

\end{appendices}

\end{document}